\pdfoutput=1 
\documentclass[preprint2]{aastex6}
\usepackage[T1]{fontenc}
\usepackage{lmodern}
\usepackage{graphicx}
\usepackage{amsmath}
\usepackage{amsfonts}
\usepackage{amssymb}

\newcommand{\figfactor}{1.0}
\newcommand{\figfactortwo}{1.0}
\newcommand{\figfactorthree}{0.6}

\begin{document}

\title{Effects of Spin on High-Energy Radiation from Accreting Black Holes}
\shorttitle{}
\author{Michael O' Riordan\altaffilmark{1}$^\dagger$, 
    Asaf Pe'er\altaffilmark{1}, 
    Jonathan C. McKinney\altaffilmark{2}}
\shortauthors{O' Riordan, Pe'er, \& McKinney}

\altaffiltext{1}{Physics Department, University College Cork, Cork, Ireland}
\altaffiltext{2}{Department of Physics and Joint Space-Science Institute,
    University of Maryland, College Park, MD 20742, USA}

\email{$^\dagger$michael\_oriordan@umail.ucc.ie}

\begin{abstract}

    Observations of jets in X-ray binaries show a correlation between radio 
    power and black hole spin.
    This correlation, if confirmed, points towards the idea that relativistic 
    jets may be powered by the rotational energy of black holes.    
    In order to examine this further,
    we perform general-relativistic radiative transport calculations on 
    magnetically arrested accretion flows, which are known to produce powerful
    jets via the Blandford-Znajek (BZ) mechanism. 
    We find that the X-ray and $\gamma$-ray emission strongly depend on spin 
    and inclination angle.
    Surprisingly, the high-energy power does not show the same dependence on
    spin as the BZ jet power, but instead can be understood as a redshift 
    effect. 
    In particular, photons observed perpendicular to the spin axis
    suffer little net redshift until originating from close to the horizon. 
    Such observers see deeper into the hot, dense, highly-magnetized inner 
    disk region.
    This effect is largest for rapidly 
    rotating black holes due to a combination of frame dragging and decreasing
    horizon radius.    
    While the X-ray emission is dominated by the near horizon region, the 
    near-infrared radiation originates at larger radii. Therefore,
    the ratio of X-ray to near-infrared power is an observational signature of
    black hole spin.

\end{abstract}
\maketitle

\section{Introduction}
\label{sec:intro}
It is widely believed that relativistic jets are powered by the rotational 
energy of black holes. 
\citet{BZ77} showed that magnetic field lines, anchored in an external accretion 
disk, are twisted by frame dragging in the vicinity of a rotating black hole.
These field lines expand 
under their own pressure, transporting energy outwards and accelerating any 
``frozen-in'' plasma into jets aligned with the spin axis.
Recent general-relativistic magnetohydrodynamic (GRMHD) simulations of 
``magnetically arrested disks'' \citep[MADs;][]{N+03} showed that 
this process can operate with efficiencies of $> 100 \%$ \citep{TNM11,MTB12}.
That is, more energy flows out of the black hole than flows in, which can only
be achieved by extracting rotational energy from the black hole.

Using 5GHz radio emission from X-ray binaries (XRBs) as a proxy for jet power, 
\citet{NM12}
found a correlation between jet power and black hole spin
\citep[but see][]{Fender+10,Russell+13}.
Their results were consistent with $P_\text{jet} \sim a^2$, 
which is the scaling derived by \citet{BZ77} for slowly rotating black holes.
They also found good agreement with the more accurate scaling 
$P_\text{jet} \sim \Omega_H^2$ \citep{TNM10,TMN12}, which works up to 
$a\approx 0.95$.
Here, $a$ is the dimensionless spin of the black hole,
$\Omega_H = a/2\,r_H$ is the angular velocity of the horizon, and
$r_H = 1 + \sqrt{1 - a^2}$ is the horizon radius 
(we work in units where $GM=c=1$, however we occasionally reintroduce factors
of $c$ for clarity).
If confirmed, this correlation provides observational evidence that jets 
are probably
powered by the rotational energy of black holes.

Although it is well established that jets produce radio emission at large radii
\citep[e.g.,][]{Fender10}, the high-energy (X-ray and $\gamma$-ray) 
radiation could originate much closer to the black hole, and so the 
contribution of jets to this radiation is less certain.
It has long been argued that inverse Compton emission from 
a corona of hot electrons surrounding the inner accretion disk can 
produce the observed X-ray spectrum in XRBs
\citep[e.g.,][]{Titarchuk94,MZ95,Gierlinski+97,Esin+97,Esin+01,Poutanen98,
    CB+06,Yuan+07,NM08,Niedzwiecki+14,Niedzwiecki+15,QL15}.
However, it is also possible that a significant fraction of the X-ray
emission originates in jets
\citep[e.g.,][]{MiRo94,Markoff+01,Markoff+03,Markoff+05,FKM04,BRP06,Kaiser06,
    GBD06,Kylafis+08,Maitra+09,PC09,PM12,Markoff+15,O'Riordan+16}.

Near the black hole where the jet originates, it is not even necessarily easy 
to distinguish what one means by a disk vs a jet due to the generically 
low plasma $\beta$ parameter and inflow-outflow regions in both the disk and 
jet \citep{MG04,McKinney06}.
Clearly, there is much uncertainty about the potentially complicated 
relationship between the high-energy emission,
the inner regions of the disk/jet, and the central black hole.
In particular, even if jets are powered by the rotational 
energy of black holes, due to the uncertainties in the source of the high-energy 
radiation, it is not clear \emph{a priori} how this radiation should
depend on spin.

To investigate this issue, we take fully three-dimensional GRMHD simulations 
with different black hole spins. We perform radiative transfer calculations 
with Comptonization to obtain the spectrum of radiation with a focus 
on high-energy radiation resolved by the region near the black hole.
We restrict our attention to the low/hard state in XRBs, since 
it is widely accepted that jets exist during this state
\citep[with transient jets launched during state transitions;][]{FBG04}.
Interestingly, although we find a strong spin dependence for the high-energy 
power, this does not follow the Blandford-Znajek (BZ) scaling.
Furthermore, the effects of spin are maximum for observers located perpendicular
to the spin axis of the black hole.
We show that the high-energy emission originates from very close to the horizon, 
and the strong spin and viewing angle dependence can be understood as a 
redshift effect.
While the X-ray power strongly depends on spin and observer inclination, the 
near-infrared (NIR) emission originates at larger radii and so is less 
sensitive to redshift effects. Therefore, for systems whose inclination angles 
are known, the ratio of X-ray to NIR power in the low/hard state can potentially 
be used to estimate spin.
Since the black hole spin does not vary between the 
low/hard and high/soft states, this ratio would compliment measurements of spin 
in the high/soft state \citep[see e.g.,][for a review]{McClintock+11}.

In Section~\ref{sec:model} we briefly describe our GRMHD simulations and
radiative transport post-processing method. 
In Section~\ref{sec:results} we show the dependence
of radiated power on spin and calculate the effects of redshift. In
Section~\ref{sec:discussion} we summarize and discuss our findings.

\section{Model}
\label{sec:model}
Radiatively inefficient accretion flows (RIAFs) have been used extensively
to model low luminosity systems such as the low/hard state in XRBs 
\citep[see e.g.,][]{NM08,YN14}. For RIAFs, by definition, the cooling time of a
fluid element is much longer than the accretion time. Therefore, radiation is 
dynamically unimportant and the evolution of the disk/jet can be calculated 
using the non-radiative GRMHD equations.
This allows a separation between the dynamical simulations
and radiative transport post-processing of the simulation results.
We use the \texttt{HARM} code \citep{Gammie+03}, which solves the GRMHD 
equations using a conservative, shock-capturing scheme.

For our purposes, we choose five MAD accretion flows with 
spins $a=\left\{0.1,\,0.2,\,0.5,\,0.9,\,0.99\right\}$ 
\citep[these are A0.1N100, A0.2N100, A0.5N100, A0.9N100, and A0.99N100 
from][]{MTB12}.
In these models, the
black hole magnetosphere compresses the inner accretion disk such that it 
becomes geometrically thin and the magneto-rotational instability is suppressed.
These MAD models efficiently extract rotational energy 
from the black hole via the  BZ mechanism, launching jets 
along the spin axis \citep{TNM11,MTB12}.
Estimates of the jet power, based on integrating fluid energy fluxes 
(dominated by the Poynting flux), 
show that the power scales as expected for the BZ mechanism
\citep[with corrections for high spins and disk thickness;][]{TMN12}.
However, such estimates are based solely on the dynamical properties of 
the fluid, and so the radiated power must be calculated in order to compare with 
observations.

Since we limit our analysis to the low/hard state in XRBs, 
we choose a black hole mass of $10M_\odot$, and accretion rate of 
$\dot{M} = 10^{-5}\dot{M}_\text{Edd}$. Such a low accretion rate ensures that 
the flow is radiatively inefficient \citep[see e.g.,][]{NM08}. 
The Eddington accretion rate, $\dot{M}_\text{Edd}$, is defined 
to be the mass accretion rate at which a disk with radiative efficiency $0.1$
would radiate at the Eddington luminosity $L_\text{Edd}$. That is,
$\dot{M}_\text{Edd}\,c^2 = 10\,L_\text{Edd}$ \citep{NM08}.
Although radio emission is expected to originate in the jet at large radii,
computational limitations force us to restrict our
analysis to the inner $r \approx 200\,r_g$, where $r_g = GM/c^2$ is the
gravitational radius. 
While the setup we use can not properly capture radio emission, 
the NIR to high-energy emission 
($\nu\gtrsim 10^{13}\,\text{Hz}$)
is dominated by regions close to the black hole,
and so setting the boundary to $r=200\,r_g$ has little effect on our results at 
these frequencies.

As discussed in \citet{O'Riordan+16}, the centre of the highly-magnetized, 
low density funnel can become artificially dense and hot due to the introduction 
of numerical density floors. We therefore remove this floor material by setting
the density to zero in regions where $b^2/\rho > \zeta$. Here, $\rho$ is the
rest mass density, and $b^2 = b^\mu b_\mu$, where $b^\mu$ is the magnetic
4-field.  The magnetic 4-field can be written in terms of the 3-field $B^i$ as
$b^\mu = h^\mu_\nu\,B^\nu / u^t$, where 
$h^\mu_\nu = \delta^\mu_\nu + u^\mu u_\nu$ is a projection tensor, $u^\mu$ is 
the fluid 4-velocity, and $B^0=0$.
We choose $\zeta = 20$ at the horizon, and linearly interpolate to $\zeta = 10$
at $r = 10$. For $r > 10$, we simply set $\zeta = 10$.
This interpolation happens to ensure that the injected floor material is 
accurately removed, without unnecessarily 
removing material very close to the black hole which can naturally become highly 
magnetized.

We calculate spectra using the same general relativistic radiative transport 
code as in \citet{O'Riordan+16}, which is based on the freely available
\texttt{grmonty} \citep{Dolence+09}. This code uses the fluid data as input, and
calculates the spectra assuming synchrotron emission, self-absorption, and 
Compton scattering from a Maxwell-J\"uttner distribution of electrons.
We assume a constant proton-to-electron temperature ratio $T_p / T_e$.
However, since differences in density and magnetization in the disk and 
jet can lead to different cooling rates for the electrons
\citep{Ressler+15,Foucart+15},
we allow this temperature ratio to vary independently in these regions.
In our models, the X-rays are dominated by the highly-magnetized inner disk 
(which is nearly indistinguishable from the jet base) and so
varying $T_p / T_e$ independently in the disk and jet has a
negligible effect on the high-energy radiation in this case. 
Therefore, we simply choose a constant ratio of $T_p/T_e = 30$
everywhere 
(we find the same dependence of the radiated power on spin with 
$T_p/T_e=3$ and $T_p/T_e=10$).

\section{Results}
\label{sec:results}
\subsection{Radiated Power}
\label{subsec:radiated_power}
\begin{figure}
    \centering
    \includegraphics[width=\figfactortwo\linewidth]{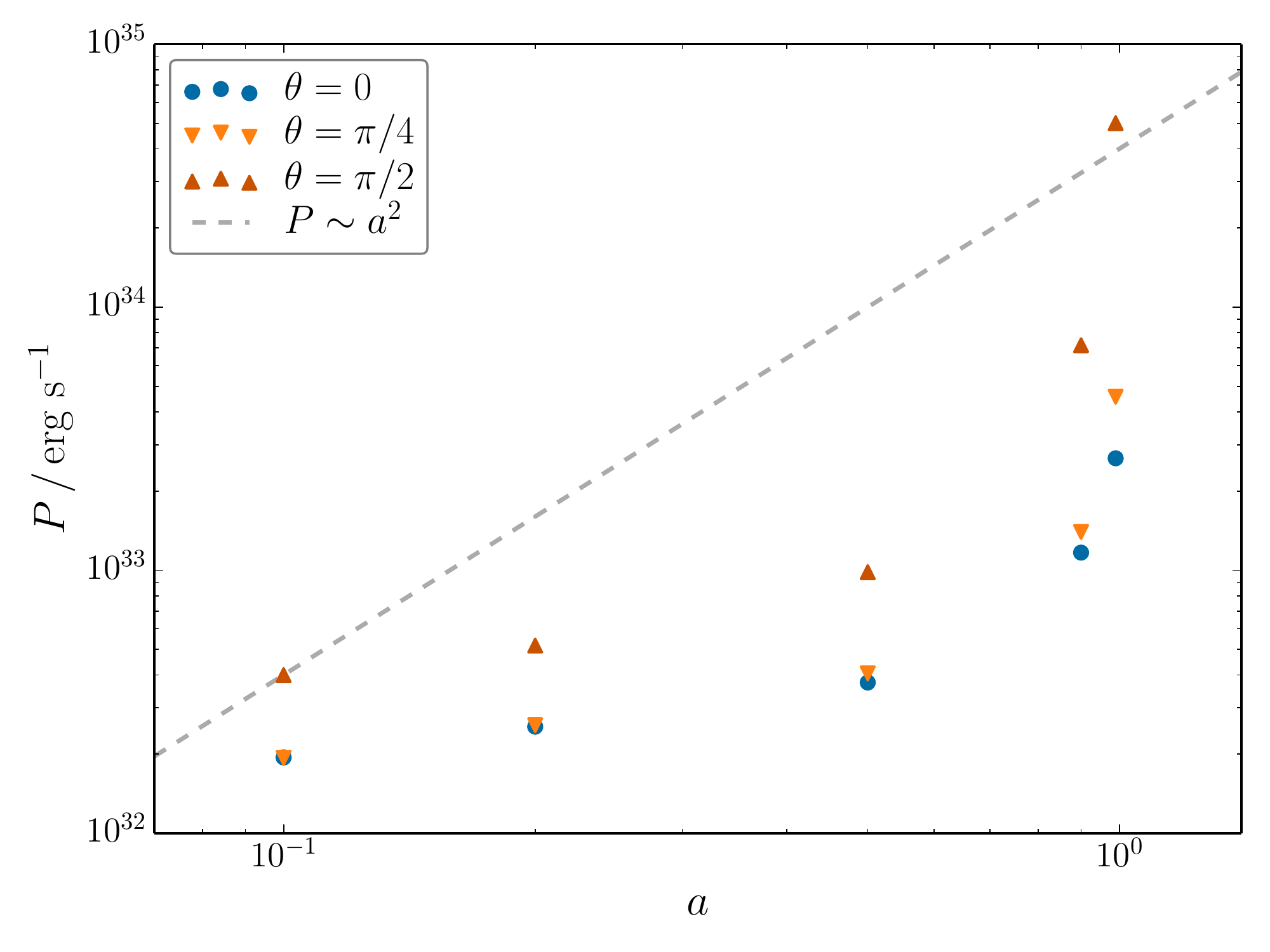}
    \caption{Integrated power vs spin for observer viewing angles of $\theta=0$
        (parallel to the spin axis), $\theta=\pi/4$,
        and $\theta=\pi/2$ 
        (perpendicular to the spin axis). The dashed line corresponds to the
        BZ scaling $P\sim a^2$.
        The dependence of the radiated power on spin clearly deviates 
        significantly from this scaling. 
        Interestingly, the effects of spin are strongest for observers
        located perpendicular to the spin axis, with a difference of more than
        two orders of magnitude in power between the $a=0.1$ and $a=0.99$ cases.
    }
    \label{fig:rad_power_vs_spin_angles}
\end{figure}
In Figure~\ref{fig:rad_power_vs_spin_angles} we show the time-averaged
radiated power 
(frequency integrated between $10^{13}$ -- $10^{24}\,\text{Hz}$)
for different spins and viewing angles. 
In what follows, any time averaging corresponds to the quasi-steady state 
between $t=10000\,r_g/c$ and $t=14000\,r_g/c$, with steps of 
$\Delta t = 400\,r_g/c$. 
We investigated the stability of this averaging
in the extreme case of $a=0.99$. Firstly, we doubled our time resolution between 
$t=10000\,r_g/c$ and $t=14000\,r_g/c$. We also increased our averaging 
window to $t=18000\,r_g/c$, and found identical results in all cases.

For observers located parallel to the 
spin axis ($\theta=0$), there is a difference of approximately one order of 
magnitude between the $a=0.1$ and $a=0.99$ cases. 
This difference increases to more than two orders of magnitude for observers 
perpendicular to the spin axis. Interestingly, the dependence of the radiated
power on spin is significantly different from the BZ scaling. 
As we show below, the origin of this discrepancy is that 
the emission in our MAD models is dominated
by the inner disk, very close to the black hole horizon. 
The strong dependence on spin and viewing angles can be understood as a 
somewhat surprising redshift effect.

\begin{figure}
    \centering
    \includegraphics[width=\figfactortwo\linewidth]{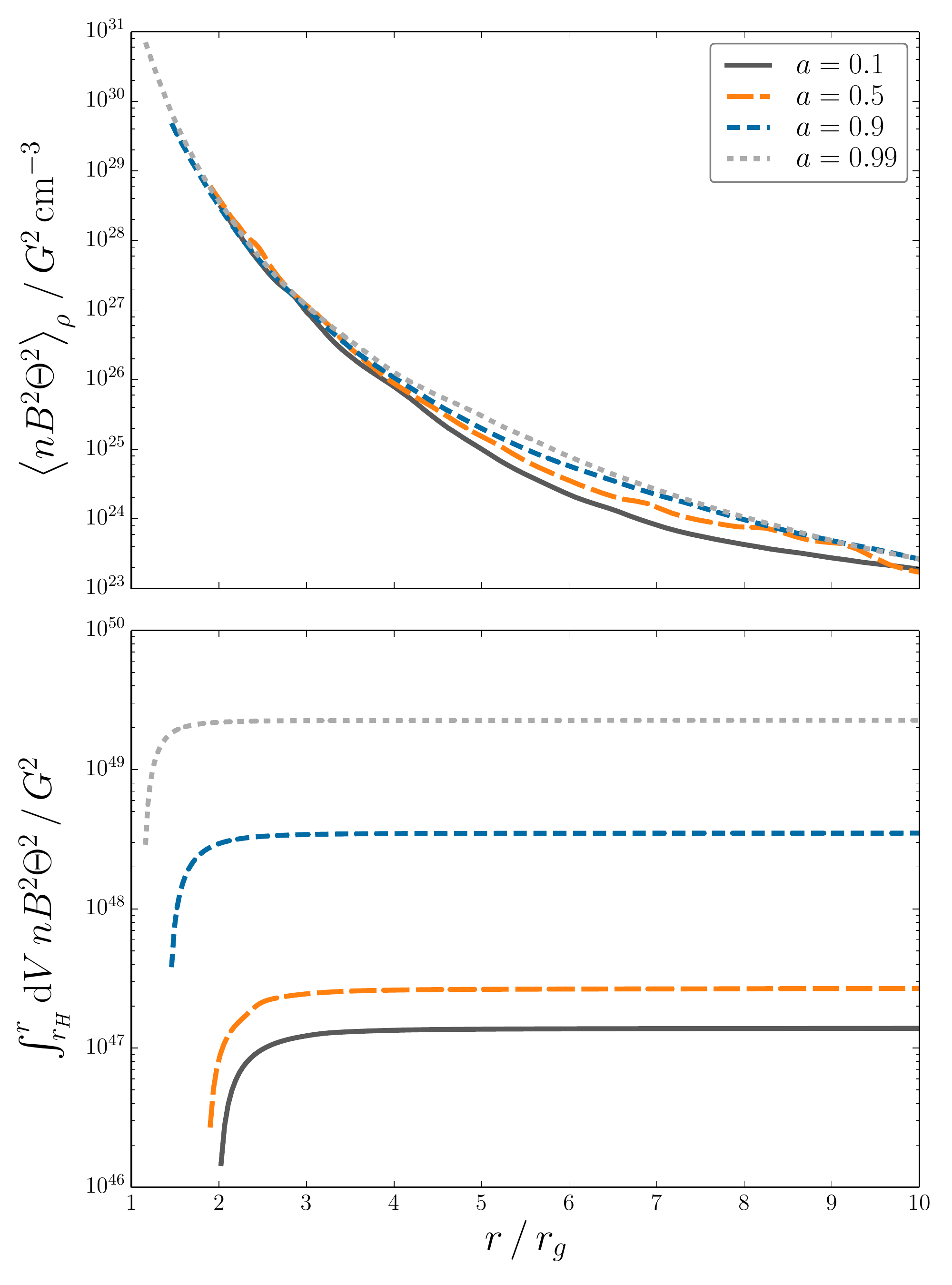}
    \caption{Time- and shell-averaged $nB^2\Theta^2$, 
        weighted by the density.
        This quantity is proportional to the total synchrotron emissivity.
        The bottom panel shows $nB^2\Theta^2$ integrated over volume
        with the integral taken between the horizon and $r$.
        This quantity is proportional to the total synchrotron power. It is 
        clear that the (comoving) radiated power is dominated by the near 
        horizon region.
    }
    \label{fig:avg_nB2T2}
\end{figure}

For any fluid quantity $Q$, we define the density-weighted, shell-average 
$\left<Q\right>_\rho$ to be
\begin{equation}
    \left<Q\right>_\rho = 
    \frac{\int\mathrm{d}A\,\rho\,Q}{\int\mathrm{d}A\,\rho}
\end{equation}
where $\mathrm{d}A=\sqrt{-g}\,\mathrm{d}x^\theta\mathrm{d}x^\phi$, and 
$g=\det\left(g_{\mu\nu}\right)$ is the metric determinant.
In all our models, the Compton $y$ parameter is $y\lesssim 1$, and so, 
to a good approximation, we can show the effects of spin on the power by 
treating synchrotron emission alone.
For a thermal electron distribution, the (comoving) synchrotron power scales as 
$P_\text{syn} \sim \int\mathrm{d}V nB^2\Theta^2$, where
$\mathrm{d}V = \sqrt{-g}\,\mathrm{d}x^r\mathrm{d}x^\theta\mathrm{d}x^\phi$,
$n$ is the electron number density, $B$ is the magnetic field strength,
and $\Theta = kT_e/mc^2$ is the dimensionless electron temperature. 
In Figure~\ref{fig:avg_nB2T2} we show the time-averaged 
$\left<nB^2\Theta^2\right>_\rho$, which is proportional to the synchrotron
emissivity $\nu j_\nu$ 
\citep[where $j_\nu$ has units of erg cm$^{-3}$ s$^{-1}$ ster$^{-1}$ Hz$^{-1}$;][]{RL79}.

Clearly, the emissivity is a strong function of $r$, and increases towards the 
black hole due to the increase in magnetic energy density and compression by the 
magnetosphere. Furthermore, the emissivity profiles are roughly independent of 
spin. This is likely a consequence of the MAD state. 
\citet{MTB12,TMN12} showed that in MAD accretion flows the magnetic flux 
saturates near the horizon, depending only weakly on spin
($\sim20\%$ difference between the $a=0$ and $a=1$ simulations). 
Therefore, for a fixed disk angular thickness, black hole mass and accretion 
rate, we expect the profile of $B^2$ to be the same for different spins. 
Close to the black hole, the fluid properties are determined by 
an approximate force balance between the inner 
magnetosphere and the thermal and ram pressures \citep{MTB12}, and so
this explains why the fluid properties are also roughly independent of spin.

The bottom panel shows the (comoving) synchrotron power
$P_\text{syn}(r)\sim\int_{r_H}^r\mathrm{d}V nB^2\Theta^2$.
It is clear from this plot that the radiated power is dominated by 
the near horizon region.
The synchrotron emissivity profiles are independent of spin, and so the
increase in power is simply a consequence of the decreasing horizon 
radius (from $r=2\,r_g$ to $r=1\,r_g$ as the spin increases from $a=0$ to $a=1$).
However, since the difference in power is due to radiation from 
$r\lesssim 2\,r_g$, it will be strongly gravitationally redshifted and so it is not 
immediately obvious that this effect is observable.
In order to check that this is in fact the reason for the spin dependence in
Figure~\ref{fig:rad_power_vs_spin_angles}, we must estimate the observed power.
That is, we must account for the effects of redshift 
(both gravitational and Doppler). 

Interestingly, as we explain in Section~\ref{subsec:redshift}, 
redshift effects naturally explain the dependence on spin and viewing angle.
In particular, for rapidly rotating black holes, frame dragging ensures that  
photons received by observers located at $\theta=\pi/2$ suffer little net 
redshift until very close to the horizon. 
In this case, there is little difference 
between the comoving and observed power, and so 
these observers see a very large increase in radiated power with spin. 
Although this effect is largest for observers 
perpendicular to the spin axis, observers located parallel to the spin
axis should also see an increase in power due to the fact that the radius
of the event horizon (i.e., the infinite redshift surface) decreases with spin.

\subsection{Redshift}
\label{subsec:redshift}
We consider the Kerr spacetime with metric $g_{\mu\nu}$ in Boyer-Lindquist 
coordinates.
We define the redshift factor to be the ratio of the energy at infinity to the
energy in the rest frame of the fluid
\begin{equation}
    \mathcal{R}=\frac{E_{\infty}}{E}=\frac{\xi^{\mu}p_{\mu}}{u^{\mu}p_{\mu}}=
    \frac{p_{t}}{u^{\mu}p_{\mu}}
\end{equation}
Here, $u^{\mu}$ is the fluid 4-velocity, $p^{\mu}$ is the photon 4-momentum,
and $\xi^\mu = \delta^\mu_t$ is the Killing vector associated with stationarity.
Since $p_{t}$ is conserved along geodesics, all the above quantities can be 
measured at the location of the emitting fluid element. 

\begin{figure}
    \centering
    \includegraphics[width=\figfactortwo\linewidth]{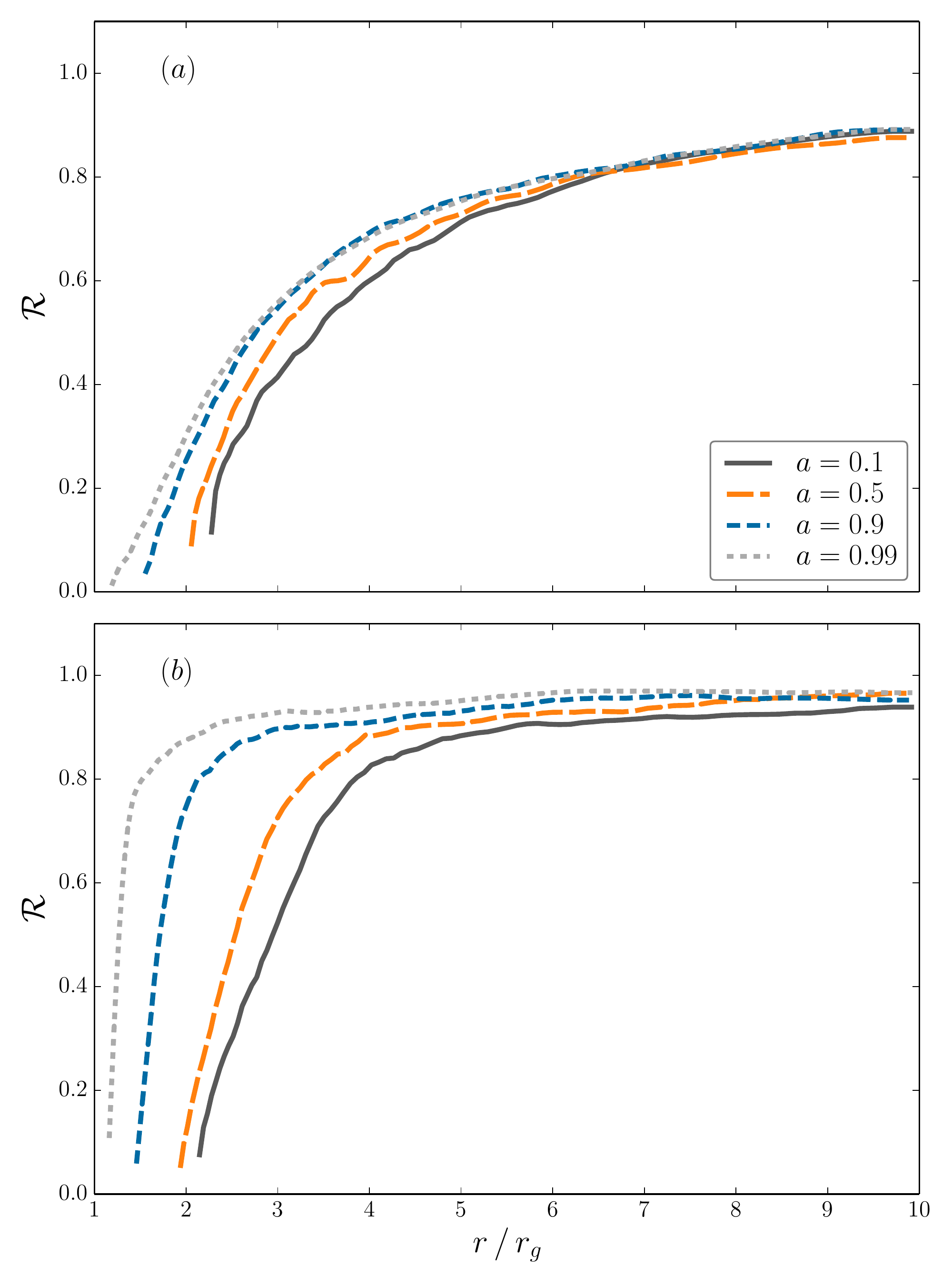}
    \caption{Redshifts for different spins and viewing 
        angles $(a):\,\theta=0$, $(b):\,\theta=\pi/2$. 
        These were calculated numerically from snapshots of the GRMHD
        data. The redshift profiles are
        much flatter for observers with $\theta=\pi/2$. That is, photons 
        received by these observers suffer little net redshift until very close
        to the horizon. Observers with $\theta=\pi/2$ see much
        deeper into the inner disk than observers with $\theta=0$.
    } 
    \label{fig:redshifts_numerical_multiple}
\end{figure}

In Figure~\ref{fig:redshifts_numerical_multiple}, we show the 
numerically calculated redshift profiles for different spins and viewing angles.
For a given spin and viewing angle, this calculation shows the average redshift 
experienced by a photon as a function of $r$.
The top panel shows the 
redshift factor for an observer with $\theta=0$. 
Close to the black hole, the $\theta$ velocity is negligible and so Doppler 
boosting is unimportant for these observers. Therefore, for $a=0.1$, 
the redshift factor is almost identical to the Schwarzschild case. 
More accurately, for observers located at $\theta=0$, the redshift is given
by the lapse
function $\alpha=\sqrt{-g_{tt} + \Omega^2g_{\phi\phi}}$, where
$\Omega=-g_{t\phi}/g_{\phi\phi}$ is the angular velocity of a 
``zero angular momentum observer'' \citep[ZAMO;][]{Bardeen+72,MT82}.
Note, however, that $\mathcal{R}=\alpha$ only if 
$\chi^\mu p_\mu=p_\phi=0$, where $\chi^\mu=\delta^\mu_\phi$ is the Killing 
vector associated with axisymmetry
(see Appendix~\ref{sec:redshift_appendix} for details).
Importantly, although these profiles are identical at large radii, they
deviate from each other close to the black hole since the horizon radius 
decreases with spin.

The bottom panel shows the redshift factor for observers with $\theta=\pi/2$.
These profiles are strikingly different from the $\theta=0$ case.
In particular, due to a combination of frame dragging and Doppler boosting, 
photons suffer little net redshift until very close to the horizon. 
Observers located at $\theta=\pi/2$ see deeper into regions of 
higher emissivity. This naturally explains the large difference in observed
power between the $\theta=0$ and $\theta=\pi/2$ inclinations.

While these calculations use model-dependent fluid data as input, 
we show in Appendix~\ref{sec:redshift_appendix} that the flattening of the 
redshift profile with spin is in fact a very general feature of rotating black 
holes. That is, the redshift profiles depend only weakly on the details of the 
accretion model, with the main contributions being black hole spin and observer 
viewing angle. Therefore, for systems in which the comoving power is dominated 
by fluid close to the horizon, we expect the high-energy emission to be a robust
signature of spin and viewing angle.

\subsection{Spectra and Observational Signatures of Black Hole Spin}
\label{subsec:spin_signatures}
\begin{figure}
    \centering
    \includegraphics[width=\figfactortwo\linewidth]{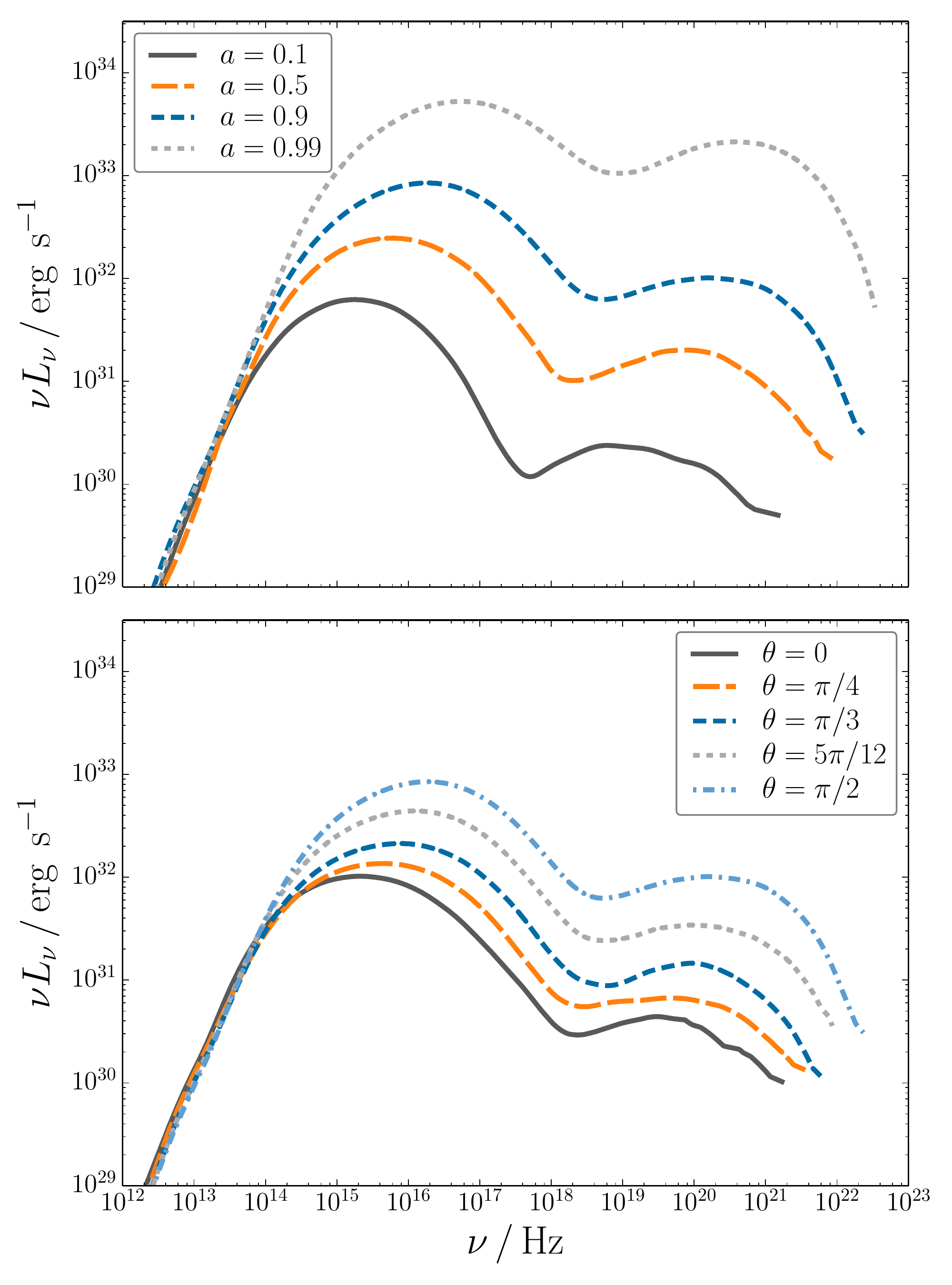}
    \caption{The top panel shows spectra for observers with $\theta=\pi/2$, 
        calculated from 
        snapshots of the fluid data. The NIR 
        emission is roughly constant with spin while the X-rays and 
        $\gamma$-rays vary significantly.
        The bottom panel shows the dependence on viewing angle for the $a=0.9$ 
        case. Both the luminosity and frequency of emission increase 
        with viewing angle.
    }
    \label{fig:spectra_spin}
\end{figure}
In Figure~\ref{fig:spectra_spin} we show spectra
for different spins and viewing angles, calculated from snapshots of the fluid 
data. 
In the top panel, we show spectra for observers with $\theta=\pi/2$, which 
maximizes the effects of spin.
For the $a=0.1$ case, the synchrotron emission peaks in the optical, while
for the $a=0.99$ case this peak increases to the X-rays.
There is also clear $\gamma$-ray emission due to inverse Compton 
scattering, which becomes more pronounced with increasing spin.
Interestingly, the NIR emission is roughly 
constant with spin while the high-energy radiation, namely the X-rays and 
$\gamma$-rays, vary significantly with spin.
The bottom panel shows the effects of varying observer inclination in the 
$a=0.9$ case. 
Both the total luminosity and frequency of the peak emission increase with 
viewing angle. Interestingly, there is little difference between the 
$\theta=0$ and $\theta=\pi/4$ inclinations 
(see also Figure~\ref{fig:rad_power_vs_spin_angles}), however the luminosity 
increases by roughly an order of magnitude between the $\theta=\pi/4$ and 
$\theta=\pi/2$ cases.

\begin{figure}
    \centering
    \includegraphics[width=\figfactortwo\linewidth]{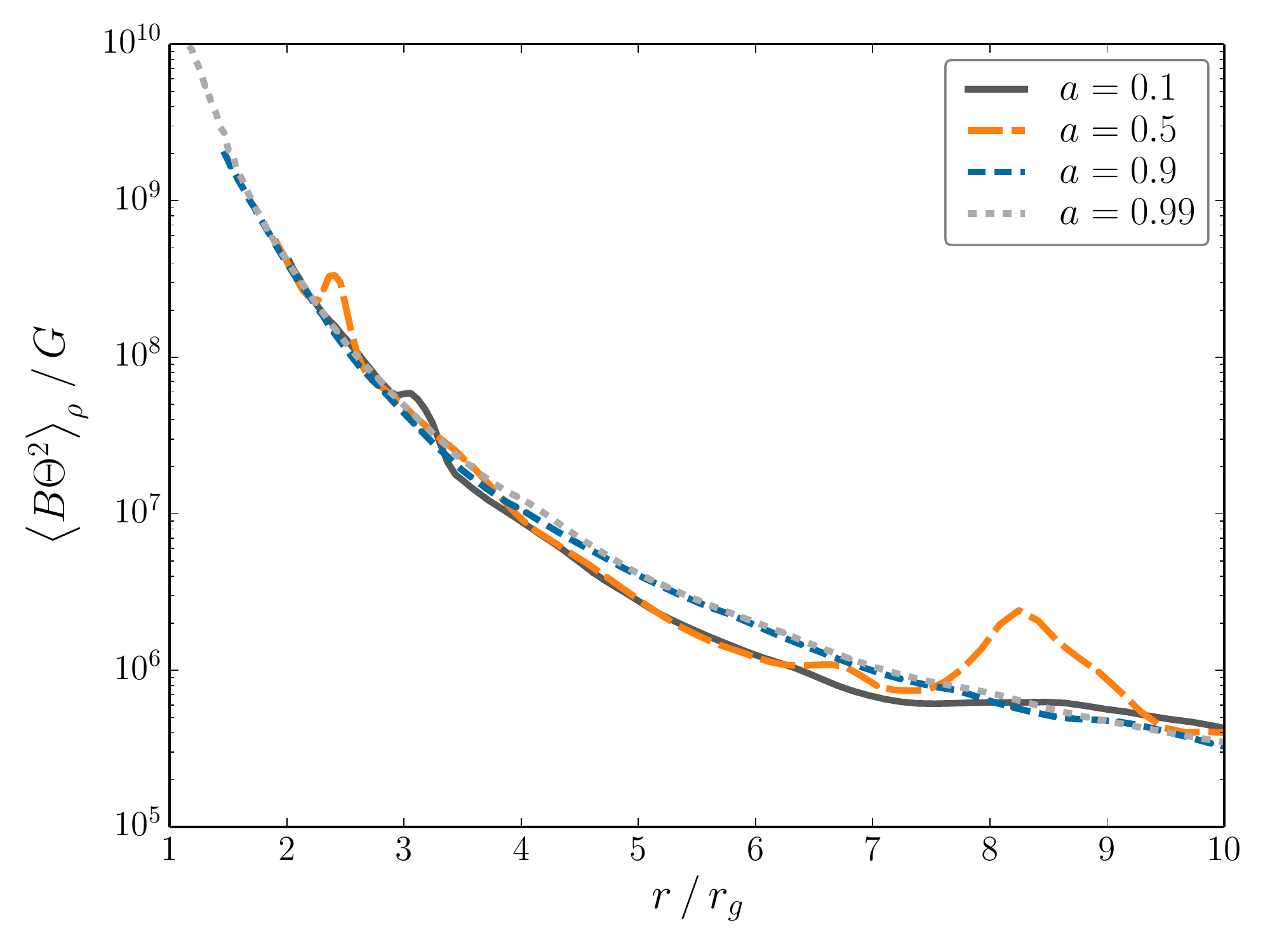}
    \caption{Time- and shell-averaged $B\Theta^2$, 
        weighted by the density.
        This quantity is proportional to the characteristic synchrotron 
        frequency.
        The high-energy emission is dominated by the near horizon region.
    }
    \label{fig:avg_BT2}
\end{figure}

As with the total radiated power, the dependence of the 
spectra on spin and viewing angle can be understood as a simple consequence of 
the emission radius.
In Figure~\ref{fig:avg_BT2} we show $\left<B\Theta^2\right>_\rho$ as a function 
of radius and spin. This quantity is proportional to the characteristic 
synchrotron frequency and so, as with the total radiated power, we expect the 
frequency of emission to increase towards the horizon. 
For higher spins and inclinations, observers receive radiation from 
smaller radii and therefore higher frequencies.
Lower frequency photons come from larger radii
and so are less sensitive to redshift effects, therefore the low-frequency power
should vary less with spin and viewing angle.
Furthermore, since the emission is dominated by the near horizon region, we 
expect the lightcurves to show significant variability over short timescales 
($\sim\text{few } r_g/c$). 
We also expect
the high-frequency emission to vary over shorter timescales than the 
low-frequency emission, with a factor of $\sim \text{few}$ difference 
between the NIR and X-ray variability timescales.

\begin{figure}
    \centering
    \includegraphics[width=\figfactortwo\linewidth]{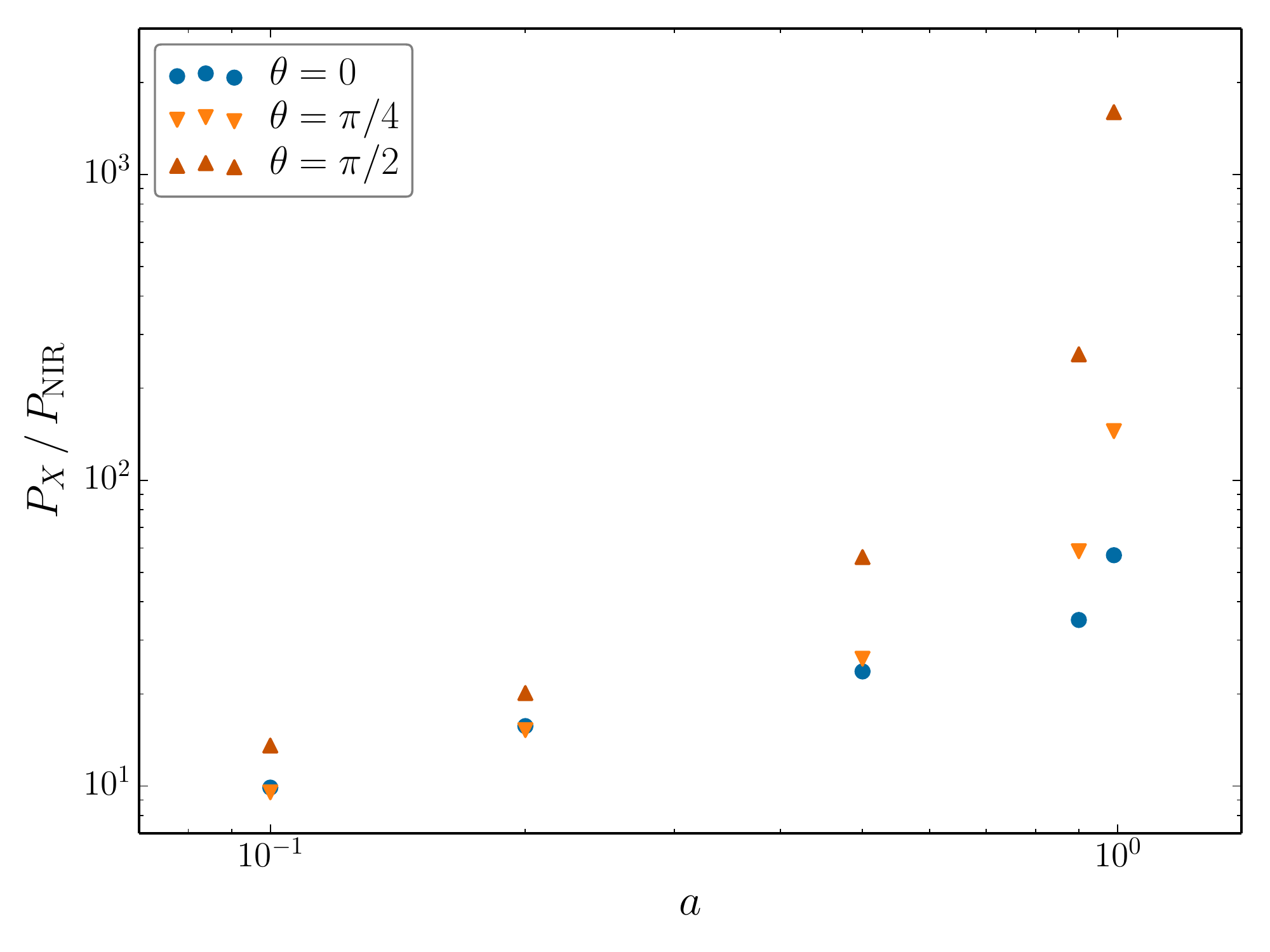}
    \caption{Ratio of the X-ray power to the NIR power for different spins and    
        viewing angles. For large inclinations, this ratio depends very strongly
        on spin.
    }
    \label{fig:PX_PNIR}
\end{figure}

In Figure~\ref{fig:PX_PNIR} we show the (time averaged) ratio of the X-ray 
(integrated between $10^{16}$ -- $10^{19}$ Hz) to NIR 
(integrated between $10^{13}$ -- $10^{14}$ Hz) power.
As expected, this ratio depends very strongly on viewing angle and spin.
Therefore, for systems whose inclination angle is known, 
especially those with large inclinations, the ratio $P_X/P_\text{NIR}$ is 
a strong signature of spin. The black hole spin likely does not vary
significantly between the low/hard and high/soft states, and so this ratio 
potentially compliments measurements of spin in the high/soft state.
Since the synchrotron frequency depends reasonably weakly on 
our choice of mass accretion rate 
($\nu_\text{syn}\sim\dot{M}^{1/2}$; see Appendix~\ref{sec:accretion_rate}), 
we expect this ratio to be a robust signature over a range of accretion rates.

\subsection{Retrograde Spin}
\label{subsec:retro}
\begin{figure}
    \centering
    \includegraphics[width=\figfactortwo\linewidth]{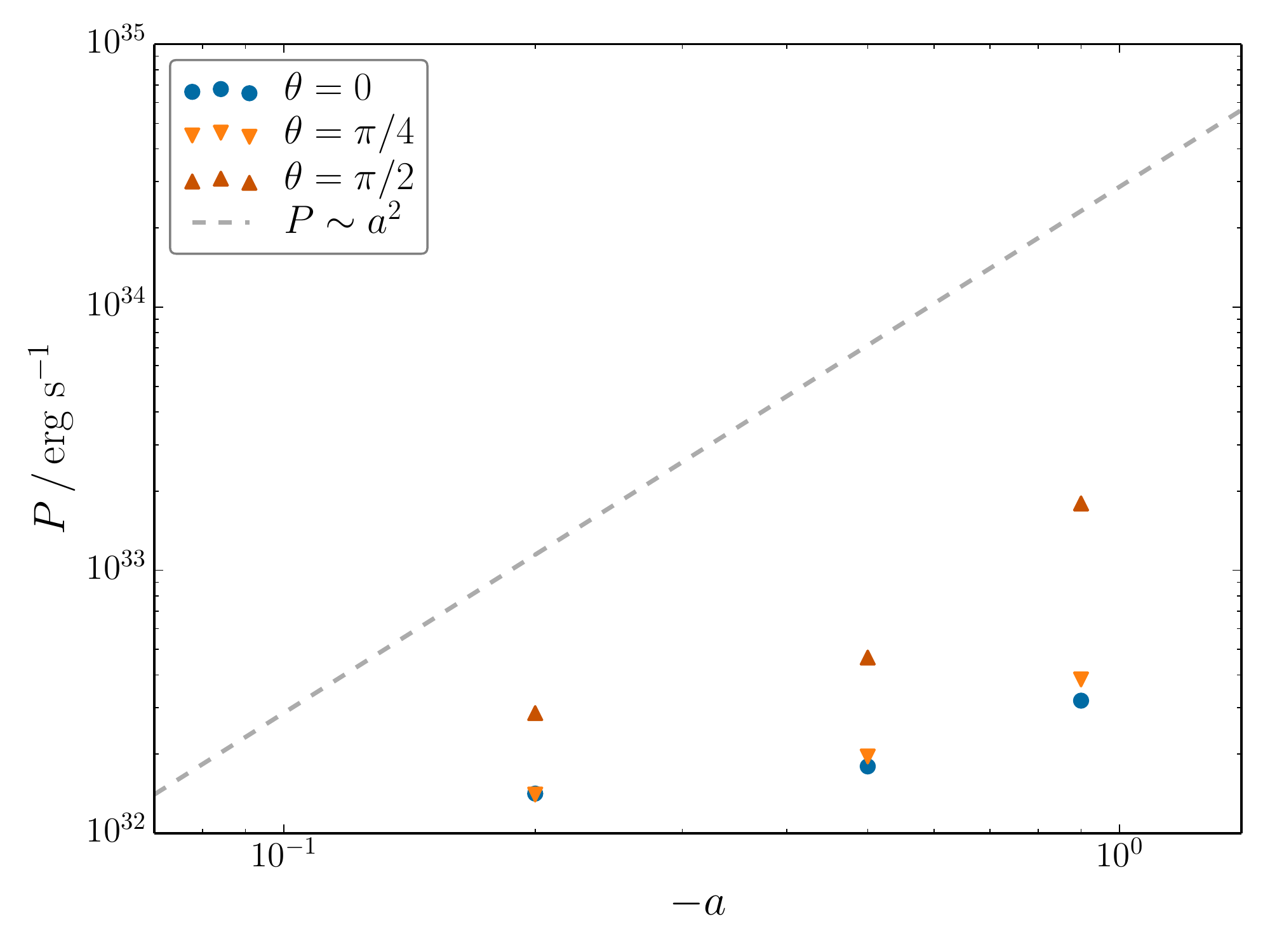}
    \caption{Integrated power vs spin for retrograde spins.        
        The total radiated power is lower in the retrograde case than the 
        prograde case due to a combination of redshift and the fact that
        less magnetic flux is trapped in the retrograde case.
    }
    \label{fig:retro}
\end{figure}

For comparison with Figure~\ref{fig:rad_power_vs_spin_angles}, in 
Figure~\ref{fig:retro} we show the integrated power vs spin for retrograde 
spins $a=\{-0.2,-0.5,-0.9\}$. As in the prograde case, the radiated power 
increases with spin and
this effect is largest for observers perpendicular to the spin axis.
Interestingly, the total radiated power is lower in the retrograde case than 
in the prograde case. 
This is likely due to the fact that prograde black holes trap 
more magnetic flux close to the horizon than retrograde black holes
\citep{TM12}. Our results show a difference of a factor of $\sim 3$ between
the $a=-0.9$ and $a=0.9$ cases, which is consistent with the findings of
\citet{TM12}. Importantly, although the radiated power is not completely 
symmetric with spin, there is clearly a degeneracy between the prograde and 
retrograde cases. 
Therefore, while the ratio of the X-ray to NIR power discussed in 
Section~\ref{subsec:spin_signatures} is an observational
probe of spin, more information would be required to distinguish between 
prograde and retrograde spins.

\subsection{Misalignment Between Jet/Disk and Spin Axis}
\label{subsec:tilt}
\begin{figure}
    \centering
    \includegraphics[width=\figfactortwo\linewidth]{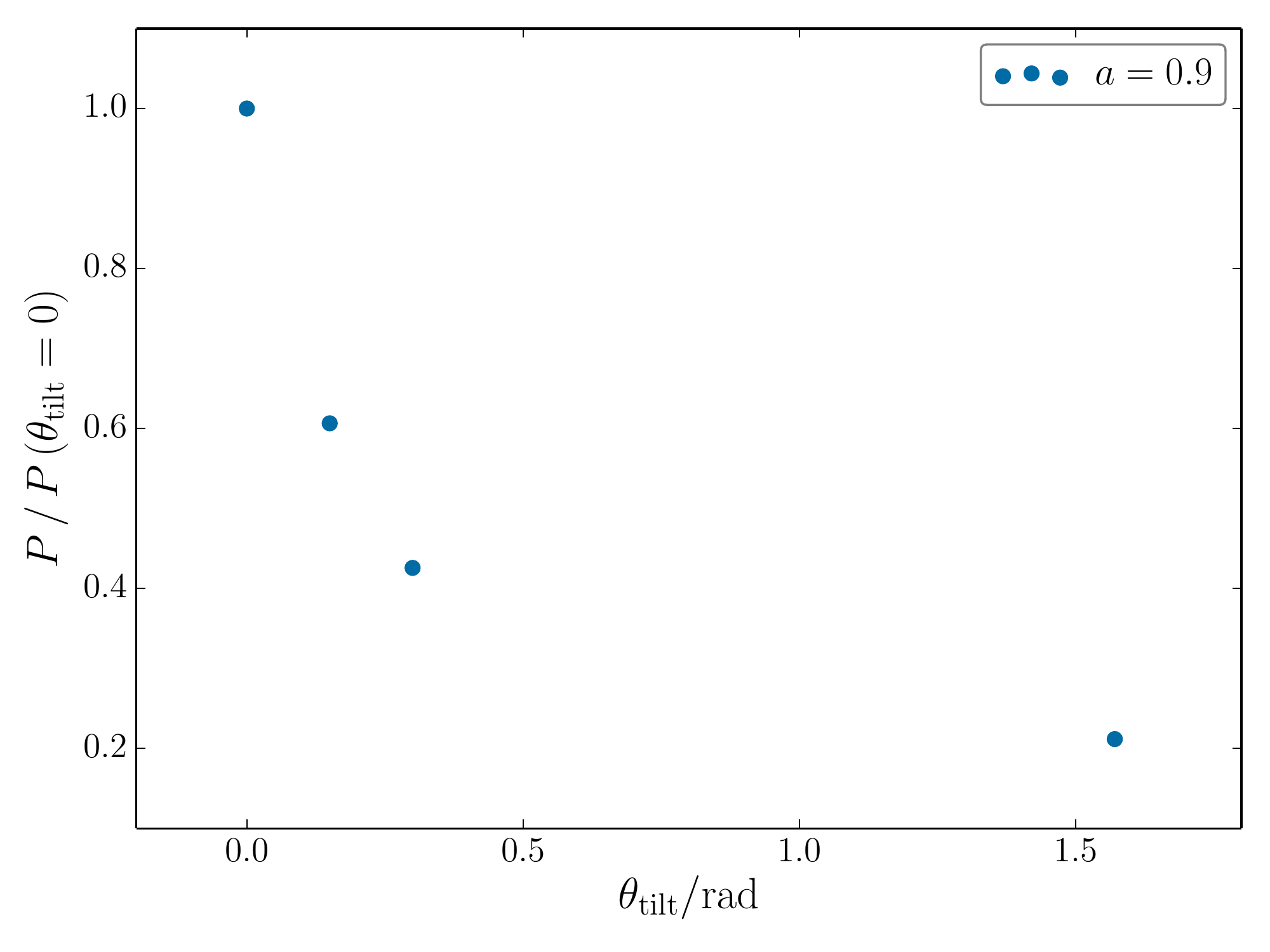}
    \caption{Observed integrated power (as a fraction of untilted power) 
        vs relative tilt angle. 
        Since the emission is dominated by the near horizon region, 
        for small misalignments there is little deviation from the 
        untilted case.
    }
    \label{fig:tilted_power}
\end{figure}
In the models considered so far, the disk angular momentum axis is aligned 
perpendicular to the black hole spin axis, while the jet points along the spin 
axis. In principle, however, the accreting plasma can have an arbitrary
angular momentum axis.
\citet{McKinney+13} studied systems in which there is a misalignment between the
disk/jet and the black hole spin axis. 
They reported a ``magneto-spin alignment'' mechanism which tends 
to align disks and jets with the rotation axis at small radii. 
Therefore, since the emission in our models is dominated by the near horizon 
region, we expect our results to be robust to minor misalignments.

In Figure~\ref{fig:tilted_power} we show the integrated power vs
relative tilt angle for the $a=0.9$ case, with the observer located 
perpendicular to the spin axis. 
The relative tilt angle, $\theta_\text{tilt}$, is defined to be the angle 
between the spin axis and the disk's angular momentum axis at large distances.
The tilt angles (in radians) are 
$\theta_\text{tilt}=\{0.0,0.15,0.3,\pi/2\}$.
There is a factor $\sim 2.5$ difference in the observed power between 
the untilted and $\theta_\text{tilt}=0.3\,\text{rad}$ 
cases, and a difference of $\sim 5$ between the untilted and fully tilted 
($\theta_\text{tilt}=\pi/2\,\text{rad}$) cases. 
We also considered a small tilt of $\theta_\text{tilt}=0.15\,\text{rad}$ 
for the extreme case of $a=0.99$, and
found a difference of $< 2$ between the untilted and tilted models. 
Therefore, we expect our results to be valid for systems with minor 
misalignments between the disk/jet and the black hole spin axis.

\section{Summary and Discussion}
\label{sec:discussion}
In this work, we calculated the effects of spin on high-energy emission 
from the low/hard state in XRBs. We modelled the low/hard state as a MAD 
accretion flow, and investigated both prograde retrograde spins. We found that
the X-ray power strongly depends on spin and observer inclination. 
In particular, the spin dependence is strikingly different from the BZ 
dependence expected for jet emission. 
In our models, the X-rays are dominated by the inner
disk, and the strong dependence on spin and viewing angle can be understood as a
redshift effect. For high spins and inclination angles, observers receive 
photons from smaller radii and therefore regions of larger synchrotron 
emissivity. Since the high-energy emission originates close to the 
horizon, it is more sensitive to spin than the low-energy emission that 
originates from larger radii. 
We identified the ratio of the X-ray power to NIR power as an
observational signature of spin. 
This quantity could potentially be used to estimate spin, and would compliment 
measurements of spin based on observations in the high/soft state.

While we expect this ratio to be particularly useful in systems with large
inclinations, in general, its dependence on quantities such as the viewing angle 
introduces significant degeneracy. Therefore, by itself, this ratio can not 
uniquely determine the black hole spin. 
However, since the high-energy spectrum in the 
low/hard state is clearly sensitive to both spin and viewing angle, it may be 
possible to use more features of the spectrum to constrain these quantities.
In particular, following the approach of the continuum-fitting (CF) and Fe line 
methods \citep[e.g.,][]{McClintock+11}, one could build up 
models of high-energy spectra for different spins and inclinations
and, for a given observational spectrum, find the best $\chi^2$ fit.
This new approach could potentially cross-validate existing methods based on 
fitting observations in the high/soft state.
A disadvantage of this method is that it can not easily distinguish between
prograde and retrograde spins. 
Both the CF and Fe line methods use the ISCO, which is monotonic with spin.
The method described here relies on the horizon radius and the effects of 
redshift and so is more symmetric with spin.
Therefore, more information would be needed to break the degeneracy between
retrograde and prograde spins.

The dependence of the high-energy power on spin is due to the 
combination of two main components, the redshift and synchrotron emissivity 
profiles. 
Interestingly, the behaviour of the redshift is in fact a very general feature 
of rotating black holes, and is largely independent of the details of accretion.
On the other hand, the emissivity itself is a model-dependent quantity. 
Our results rely on the fact that the comoving synchrotron power in our MAD 
models is strongly dominated by the near horizon region. The observed 
high-energy radiation should therefore be highly variable on timescales
of the order of a few light crossing times. Furthermore, we expect the 
variability timescale for the lower frequency emission to be longer since this
originates at larger radii.

The spectra shown in Figure~\ref{fig:spectra_spin} are consistent with the
basic 
observed X-ray hardness/flux relations for XRBs in the low/hard state 
\citep{FBG04}.
The time-averaged X-ray hardness ratio 
\citep[defined to be the ratio of the flux at 
$6.3$--$10.5$ keV to the flux at $3.8$--$7.3$ keV;][]{FBG04,Belloni+05} 
varies between 0.7 and 0.9, with higher spins slightly softer than lower spins.
The luminosities in the low spin cases are likely 
somewhat lower than expected for the low/hard state, and are probably more 
consistent with the so-called quiescent state \citep[e.g.,][]{RM06}.
However, this is not a serious issue. As we show in 
Appendix~\ref{sec:accretion_rate}, small changes in the accretion rate can
significantly increase the total luminosity without greatly affecting the 
frequency of emission. 
Therefore, increasing the luminosity would not change our conclusions 
regarding the scaling in Figure~\ref{fig:rad_power_vs_spin_angles} or
the ratio $P_X/P_\text{NIR}$ in Figure~\ref{fig:PX_PNIR}.

\citet{Moscibrodzka+09} considered the effects of spin and viewing angle on
radiation from non-MAD \citep[called SANE in ][]{Narayan+12} 
accretion flows in the context of Sgr A*. 
Interestingly, while they found that the X-ray flux increases dramatically with 
both spin and observer inclination, they attribute this dependence to a 
different effect than the one described here.
In their models, the X-ray emission is produced by scattering from hot 
electrons at $r=r_\text{ISCO}$, and so the dependence on spin manifests itself in
a very similar manner to thin disks \citep[see e.g.,][]{McClintock+11}.
In our models, by contrast, most of the observed high-energy radiation 
originates from right outside the horizon, with the ISCO playing no special 
role.
This can likely be attributed to the fact that the disks considered here are
geometrically thicker, and so the density does not drop off significantly inside 
the ISCO. 
Therefore, the our results are probably more relevant for low 
luminosity, radiatively inefficient systems, in which the disk is expected to be
geometrically thick.

Furthermore, our work improves upon this study in two major areas.
Firstly, our simulations are fully 3D, which is required to avoid decaying 
turbulence and reach a well defined steady state \citep{Cowling33,Sadowski+15}.
Axisymmetric simulations can not reliably capture the effects of spin, 
since the resulting radiation will be influenced by the extent to which the 
spin has affected the flow by the time the turbulence decays.
Secondly, in MAD models, the final amount of magnetic flux at the horizon is 
independent of the initial flux content of the torus, which in SANE models can 
artificially introduce a spin dependence \citep{TM12, TMN12}. 
Therefore, MAD models are more reliable for studying the effects of spin on the
high-energy radiation.

While our calculations apply to MAD accretion flows in the 
low/hard state, the redshift effects described here might also be important when
considering thin MADs in the high/soft state. 
\citet{avara+15} demonstrated an $80\%$ deviation from
the standard Novikov-Thorne radiative efficiency, with most of the radiation
coming from at or below the ISCO.
As shown here, for rapidly spinning black holes, radiation from small radii is 
very strongly affected by variations in spin and viewing angle. Therefore,
if the radiation from thin MADs originates at smaller radii than expected 
for standard thin disks, this could have important implications for measurements
of spin in the high/soft state.

Our analysis was carried out for a black hole mass of $M=10M_\odot$.
However, since the relevant length and time scales are set by $M$, 
we can scale our results to arbitrary masses as follows. 
Assuming that the accretion rate is a fixed fraction of the 
Eddington rate $\dot{M} \sim \dot{M}_\text{Edd} \sim M$, from
Appendix~\ref{sec:accretion_rate} we find that
$n \sim M^{-1}$, $B \sim M^{-1/2}$, and $\Theta \sim M^0$.
These relationships can be used to scale the spectral 
features in Figure~\ref{fig:spectra_spin} to supermassive black holes. 
Importantly, however, this scaling is only appropriate for systems which are 
well described by RIAFs. Therefore, our results are potentially relevant for
accreting supermassive black hole systems such as Sgr A* and low luminosity 
subclasses of AGN such as LINERS and BL Lac objects \citep[see e.g.,][]{YN14}.
Although BL Lacs (and blazars in general) have jets roughly 
aligned with the observer, at small radii there could be a misalignment between
the jet and spin axes (see Section~\ref{subsec:tilt}).
Such a misalignment could significantly enhance the high-energy emission 
from close to the black hole, leading to the intriguing possibility that
near-horizon emission is responsible for the short-timescale variability
observed in these systems 
\citep[e.g.,][]{Aharonian+07,Albert+07,Aleksic+11,Ackermann+16}.

The current work is somewhat limited by the assumption of a thermal distribution 
of electrons. The highly-magnetized inner disk region could contain a significant
number of non-thermal particles due to acceleration by magnetic reconnection
\citep[e.g.,][]{SS14}. 
However, thermal electrons might dominate emission from near the horizon, 
as has been sufficient to explain the low-hard like state in Sgr A* and M87
\citep{Dexter+12,Broderick+14,BT15}.
Furthermore, different prescriptions for treating the electron temperature might 
reduce the dominance of emission from the inner disk and instead
``light up'' the funnel wall region \citep[e.g.,][]{MF13,Moscibrodzka+14}.
These prescriptions usually separate the jet and disk based on $b^2/\rho$ or 
the plasma $\beta$. In our models, the inner disk is
highly magnetized and so differentiating between the jet and disk based on the 
magnetization alone would in fact treat the inner disk region in a similar 
manner to the jet. 
The treatment of the electron physics in accretion disks and jets remains an 
active area of research, and we will apply our results with new models of 
electron physics as they become available.

\acknowledgments 
The authors would like to thank the anonymous referee
for helpful suggestions that have improved the quality of the manuscript.
JCM would like to thank Alexander Tchekhovskoy for providing simulation data.
MOR is supported by the Irish Research Council under grant number GOIPG/2013/315.
This research was partially supported by the European Union Seventh Framework 
Programme (FP7/2007-2013) under grant agreement no 618499.
JCM acknowledges NASA/NSF/TCAN (NNX14AB46G), NSF/XSEDE/TACC (TGPHY120005), and 
NASA/Pleiades (SMD-14-5451).



\bibliographystyle{hapj}
\bibliography{rad_spin}

\begin{thebibliography}{70}
\expandafter\ifx\csname natexlab\endcsname\relax\def\natexlab#1{#1}\fi

\bibitem[{{Ackermann} {et~al.}(2016){Ackermann}, {Anantua}, {Asano}, {Baldini},
  {Barbiellini}, {Bastieri}, {Becerra Gonzalez}, {Bellazzini}, {Bissaldi},
  {Blandford}, {Bloom}, {Bonino}, {Bottacini}, {Bruel}, {Buehler}, {Caliandro},
  {Cameron}, {Caragiulo}, {Caraveo}, {Cavazzuti}, {Cecchi}, {Cheung}, {Chiang},
  {Chiaro}, {Ciprini}, {Cohen-Tanugi}, {Costanza}, {Cutini}, {D'Ammando}, {de
  Palma}, {Desiante}, {Digel}, {Di Lalla}, {Di Mauro}, {Di Venere}, {Drell},
  {Favuzzi}, {Fegan}, {Ferrara}, {Fukazawa}, {Funk}, {Fusco}, {Gargano},
  {Gasparrini}, {Giglietto}, {Giordano}, {Giroletti}, {Grenier}, {Guillemot},
  {Guiriec}, {Hayashida}, {Hays}, {Horan}, {J{\'o}hannesson}, {Kensei},
  {Kocevski}, {Kuss}, {La Mura}, {Larsson}, {Latronico}, {Li}, {Longo},
  {Loparco}, {Lott}, {Lovellette}, {Lubrano}, {Madejski}, {Magill}, {Maldera},
  {Manfreda}, {Mayer}, {Mazziotta}, {Michelson}, {Mirabal}, {Mizuno},
  {Monzani}, {Morselli}, {Moskalenko}, {Nalewajko}, {Negro}, {Nuss}, {Ohsugi},
  {Orlando}, {Paneque}, {Perkins}, {Pesce-Rollins}, {Piron}, {Pivato},
  {Porter}, {Principe}, {Rando}, {Razzano}, {Razzaque}, {Reimer}, {Scargle},
  {Sgr{\`o}}, {Sikora}, {Simone}, {Siskind}, {Spada}, {Spinelli}, {Stawarz},
  {Thayer}, {Thompson}, {Torres}, {Troja}, {Uchiyama}, {Yuan}, \&
  {Zimmer}}]{Ackermann+16}
{Ackermann}, M. {et~al.} 2016, \apjl, 824, L20, 1605.05324

\bibitem[{{Aharonian} {et~al.}(2007){Aharonian}, {Akhperjanian}, {Bazer-Bachi},
  {Behera}, {Beilicke}, {Benbow}, {Berge}, {Bernl{\"o}hr}, \& et.
  al.}]{Aharonian+07}
{Aharonian}, F. {et~al.} 2007, \apjl, 664, L71, 0706.0797

\bibitem[{{Albert} {et~al.}(2007){Albert}, {Aliu}, {Anderhub}, {Antoranz},
  {Armada}, {Baixeras}, {Barrio}, {Bartko}, \& et. al.}]{Albert+07}
{Albert}, J. {et~al.} 2007, \apj, 669, 862, arXiv:astro-ph/0702008

\bibitem[{{Aleksi{\'c}} {et~al.}(2011){Aleksi{\'c}}, {Antonelli}, {Antoranz},
  {Backes}, {Barrio}, {Bastieri}, {Becerra Gonz{\'a}lez}, {Bednarek}, \& et.
  al.}]{Aleksic+11}
{Aleksi{\'c}}, J. {et~al.} 2011, \apjl, 730, L8, 1101.4645

\bibitem[{{Avara} {et~al.}(2015){Avara}, {McKinney}, \& {Reynolds}}]{avara+15}
{Avara}, M.~J., {McKinney}, J.~C., \& {Reynolds}, C.~S. 2015, ArXiv e-prints,
  1508.05323

\bibitem[{{Bardeen} {et~al.}(1972){Bardeen}, {Press}, \&
  {Teukolsky}}]{Bardeen+72}
{Bardeen}, J.~M., {Press}, W.~H., \& {Teukolsky}, S.~A. 1972, \apj, 178, 347

\bibitem[{{Belloni} {et~al.}(2005){Belloni}, {Homan}, {Casella}, {van der
  Klis}, {Nespoli}, {Lewin}, {Miller}, \& {M{\'e}ndez}}]{Belloni+05}
{Belloni}, T., {Homan}, J., {Casella}, P., {van der Klis}, M., {Nespoli}, E.,
  {Lewin}, W.~H.~G., {Miller}, J.~M., \& {M{\'e}ndez}, M. 2005, \aap, 440, 207,
  astro-ph/0504577

\bibitem[{{Blandford} \& {Znajek}(1977)}]{BZ77}
{Blandford}, R.~D., \& {Znajek}, R.~L. 1977, \mnras, 179, 433

\bibitem[{{Bosch-Ramon} {et~al.}(2006){Bosch-Ramon}, {Romero}, \&
  {Paredes}}]{BRP06}
{Bosch-Ramon}, V., {Romero}, G.~E., \& {Paredes}, J.~M. 2006, \aap, 447, 263,
  arXiv:astro-ph/0509086

\bibitem[{{Broderick} {et~al.}(2014){Broderick}, {Johannsen}, {Loeb}, \&
  {Psaltis}}]{Broderick+14}
{Broderick}, A.~E., {Johannsen}, T., {Loeb}, A., \& {Psaltis}, D. 2014, \apj,
  784, 7, 1311.5564

\bibitem[{{Broderick} \& {Tchekhovskoy}(2015)}]{BT15}
{Broderick}, A.~E., \& {Tchekhovskoy}, A. 2015, \apj, 809, 97, 1506.04754

\bibitem[{{Cadolle Bel} {et~al.}(2006){Cadolle Bel}, {Sizun}, {Goldwurm},
  {Rodriguez}, {Laurent}, {Zdziarski}, {Foschini}, {Goldoni}, {Gouiff{\`e}s},
  {Malzac}, {Jourdain}, \& {Roques}}]{CB+06}
{Cadolle Bel}, M. {et~al.} 2006, \aap, 446, 591, arXiv:astro-ph/0509851

\bibitem[{{Cowling}(1933)}]{Cowling33}
{Cowling}, T.~G. 1933, \mnras, 94, 39

\bibitem[{{Cunningham} \& {Bardeen}(1972)}]{CB72}
{Cunningham}, C.~T., \& {Bardeen}, J.~M. 1972, \apjl, 173, L137

\bibitem[{{Cunningham} \& {Bardeen}(1973)}]{CB73}
------. 1973, \apj, 183, 237

\bibitem[{{Dexter} {et~al.}(2012){Dexter}, {McKinney}, \& {Agol}}]{Dexter+12}
{Dexter}, J., {McKinney}, J.~C., \& {Agol}, E. 2012, \mnras, 421, 1517,
  1109.6011

\bibitem[{{Dolence} {et~al.}(2009){Dolence}, {Gammie}, {Mo{\'s}cibrodzka}, \&
  {Leung}}]{Dolence+09}
{Dolence}, J.~C., {Gammie}, C.~F., {Mo{\'s}cibrodzka}, M., \& {Leung}, P.~K.
  2009, \apjs, 184, 387

\bibitem[{{Esin} {et~al.}(2001){Esin}, {McClintock}, {Drake}, {Garcia},
  {Haswell}, {Hynes}, \& {Muno}}]{Esin+01}
{Esin}, A.~A., {McClintock}, J.~E., {Drake}, J.~J., {Garcia}, M.~R., {Haswell},
  C.~A., {Hynes}, R.~I., \& {Muno}, M.~P. 2001, \apj, 555, 483,
  arXiv:astro-ph/0103044

\bibitem[{{Esin} {et~al.}(1997){Esin}, {McClintock}, \& {Narayan}}]{Esin+97}
{Esin}, A.~A., {McClintock}, J.~E., \& {Narayan}, R. 1997, \apj, 489, 865,
  arXiv:astro-ph/9705237

\bibitem[{{Falcke} {et~al.}(2004){Falcke}, {K{\"o}rding}, \& {Markoff}}]{FKM04}
{Falcke}, H., {K{\"o}rding}, E., \& {Markoff}, S. 2004, \aap, 414, 895,
  arXiv:astro-ph/0305335

\bibitem[{{Fanton} {et~al.}(1997){Fanton}, {Calvani}, {de Felice}, \&
  {Cadez}}]{Fanton+97}
{Fanton}, C., {Calvani}, M., {de Felice}, F., \& {Cadez}, A. 1997, \pasj, 49,
  159

\bibitem[{{Fender}(2010)}]{Fender10}
{Fender}, R. 2010, in Lecture Notes in Physics, Berlin Springer Verlag, Vol.
  794, Lecture Notes in Physics, Berlin Springer Verlag, ed. T.~{Belloni}, 115,
  0909.2572

\bibitem[{{Fender} {et~al.}(2004){Fender}, {Belloni}, \& {Gallo}}]{FBG04}
{Fender}, R.~P., {Belloni}, T.~M., \& {Gallo}, E. 2004, \mnras, 355, 1105,
  arXiv:astro-ph/0409360

\bibitem[{{Fender} {et~al.}(2010){Fender}, {Gallo}, \& {Russell}}]{Fender+10}
{Fender}, R.~P., {Gallo}, E., \& {Russell}, D. 2010, \mnras, 406, 1425,
  1003.5516

\bibitem[{{Foucart} {et~al.}(2015){Foucart}, {Chandra}, {Gammie}, \&
  {Quataert}}]{Foucart+15}
{Foucart}, F., {Chandra}, M., {Gammie}, C.~F., \& {Quataert}, E. 2015, ArXiv
  e-prints, 1511.04445

\bibitem[{{Gammie} {et~al.}(2003){Gammie}, {McKinney}, \&
  {T{\'o}th}}]{Gammie+03}
{Gammie}, C.~F., {McKinney}, J.~C., \& {T{\'o}th}, G. 2003, \apj, 589, 444,
  arXiv:astro-ph/0301509

\bibitem[{{Gierlinski} {et~al.}(1997){Gierlinski}, {Zdziarski}, {Done},
  {Johnson}, {Ebisawa}, {Ueda}, {Haardt}, \& {Phlips}}]{Gierlinski+97}
{Gierlinski}, M., {Zdziarski}, A.~A., {Done}, C., {Johnson}, W.~N., {Ebisawa},
  K., {Ueda}, Y., {Haardt}, F., \& {Phlips}, B.~F. 1997, \mnras, 288, 958,
  arXiv:astro-ph/9610156

\bibitem[{{Gupta} {et~al.}(2006){Gupta}, {B{\"o}ttcher}, \& {Dermer}}]{GBD06}
{Gupta}, S., {B{\"o}ttcher}, M., \& {Dermer}, C.~D. 2006, \apj, 644, 409,
  arXiv:astro-ph/0602439

\bibitem[{{Kaiser}(2006)}]{Kaiser06}
{Kaiser}, C.~R. 2006, \mnras, 367, 1083, arXiv:astro-ph/0601103

\bibitem[{{Kylafis} {et~al.}(2008){Kylafis}, {Papadakis}, {Reig}, {Giannios},
  \& {Pooley}}]{Kylafis+08}
{Kylafis}, N.~D., {Papadakis}, I.~E., {Reig}, P., {Giannios}, D., \& {Pooley},
  G.~G. 2008, \aap, 489, 481, 0807.2910

\bibitem[{{MacDonald} \& {Thorne}(1982)}]{MT82}
{MacDonald}, D., \& {Thorne}, K.~S. 1982, \mnras, 198, 345

\bibitem[{{Magdziarz} \& {Zdziarski}(1995)}]{MZ95}
{Magdziarz}, P., \& {Zdziarski}, A.~A. 1995, \mnras, 273, 837

\bibitem[{{Maitra} {et~al.}(2009){Maitra}, {Markoff}, {Brocksopp}, {Noble},
  {Nowak}, \& {Wilms}}]{Maitra+09}
{Maitra}, D., {Markoff}, S., {Brocksopp}, C., {Noble}, M., {Nowak}, M., \&
  {Wilms}, J. 2009, \mnras, 398, 1638, 0904.2128

\bibitem[{{Markoff} {et~al.}(2001){Markoff}, {Falcke}, \&
  {Fender}}]{Markoff+01}
{Markoff}, S., {Falcke}, H., \& {Fender}, R. 2001, \aap, 372, L25,
  arXiv:astro-ph/0010560

\bibitem[{{Markoff} {et~al.}(2003){Markoff}, {Nowak}, {Corbel}, {Fender}, \&
  {Falcke}}]{Markoff+03}
{Markoff}, S., {Nowak}, M., {Corbel}, S., {Fender}, R., \& {Falcke}, H. 2003,
  \aap, 397, 645, arXiv:astro-ph/0210439

\bibitem[{{Markoff} {et~al.}(2015){Markoff}, {Nowak}, {Gallo}, {Hynes},
  {Wilms}, {Plotkin}, {Maitra}, {Silva}, \& {Drappeau}}]{Markoff+15}
{Markoff}, S. {et~al.} 2015, ArXiv e-prints, 1510.02244

\bibitem[{{Markoff} {et~al.}(2005){Markoff}, {Nowak}, \& {Wilms}}]{Markoff+05}
{Markoff}, S., {Nowak}, M.~A., \& {Wilms}, J. 2005, \apj, 635, 1203,
  arXiv:astro-ph/0509028

\bibitem[{{McClintock} {et~al.}(2011){McClintock}, {Narayan}, {Davis}, {Gou},
  {Kulkarni}, {Orosz}, {Penna}, {Remillard}, \& {Steiner}}]{McClintock+11}
{McClintock}, J.~E. {et~al.} 2011, Classical and Quantum Gravity, 28, 114009

\bibitem[{{McKinney}(2006)}]{McKinney06}
{McKinney}, J.~C. 2006, \mnras, 368, 1561, arXiv:astro-ph/0603045

\bibitem[{{McKinney} \& {Gammie}(2004)}]{MG04}
{McKinney}, J.~C., \& {Gammie}, C.~F. 2004, \apj, 611, 977,
  arXiv:astro-ph/0404512

\bibitem[{{McKinney} {et~al.}(2012){McKinney}, {Tchekhovskoy}, \&
  {Blandford}}]{MTB12}
{McKinney}, J.~C., {Tchekhovskoy}, A., \& {Blandford}, R.~D. 2012, \mnras, 423,
  3083, 1201.4163

\bibitem[{{McKinney} {et~al.}(2013){McKinney}, {Tchekhovskoy}, \&
  {Blandford}}]{McKinney+13}
------. 2013, Science, 339, 49, 1211.3651

\bibitem[{{Mirabel} \& {Rodr{\'{\i}}guez}(1994)}]{MiRo94}
{Mirabel}, I.~F., \& {Rodr{\'{\i}}guez}, L.~F. 1994, \nat, 371, 46

\bibitem[{{Mo{\'s}cibrodzka} \& {Falcke}(2013)}]{MF13}
{Mo{\'s}cibrodzka}, M., \& {Falcke}, H. 2013, \aap, 559, L3

\bibitem[{{Mo{\'s}cibrodzka} {et~al.}(2014){Mo{\'s}cibrodzka}, {Falcke},
  {Shiokawa}, \& {Gammie}}]{Moscibrodzka+14}
{Mo{\'s}cibrodzka}, M., {Falcke}, H., {Shiokawa}, H., \& {Gammie}, C.~F. 2014,
  \aap, 570, A7

\bibitem[{{Mo{\'s}cibrodzka} {et~al.}(2009){Mo{\'s}cibrodzka}, {Gammie},
  {Dolence}, {Shiokawa}, \& {Leung}}]{Moscibrodzka+09}
{Mo{\'s}cibrodzka}, M., {Gammie}, C.~F., {Dolence}, J.~C., {Shiokawa}, H., \&
  {Leung}, P.~K. 2009, \apj, 706, 497

\bibitem[{{Narayan} {et~al.}(2003){Narayan}, {Igumenshchev}, \&
  {Abramowicz}}]{N+03}
{Narayan}, R., {Igumenshchev}, I.~V., \& {Abramowicz}, M.~A. 2003, \pasj, 55,
  L69, astro-ph/0305029

\bibitem[{{Narayan} \& {McClintock}(2008)}]{NM08}
{Narayan}, R., \& {McClintock}, J.~E. 2008, NewAR, 51, 733, 0803.0322

\bibitem[{{Narayan} \& {McClintock}(2012)}]{NM12}
------. 2012, \mnras, 419, L69, 1112.0569

\bibitem[{{Narayan} {et~al.}(2012){Narayan}, {Sadowski}, {Penna}, \&
  {Kulkarni}}]{Narayan+12}
{Narayan}, R., {Sadowski}, A., {Penna}, R.~F., \& {Kulkarni}, A.~K. 2012,
  \mnras, 426, 3241, 1206.1213

\bibitem[{{Nied{\'z}wiecki} {et~al.}(2014){Nied{\'z}wiecki}, {Xie}, \& {St{\c
  e}pnik}}]{Niedzwiecki+15}
{Nied{\'z}wiecki}, A., {Xie}, F.-G., \& {St{\c e}pnik}, A. 2014, \mnras, 443,
  1733, 1406.6003

\bibitem[{{Nied{\'z}wiecki} {et~al.}(2012){Nied{\'z}wiecki}, {Xie}, \&
  {Zdziarski}}]{Niedzwiecki+14}
{Nied{\'z}wiecki}, A., {Xie}, F.-G., \& {Zdziarski}, A.~A. 2012, \mnras, 420,
  1195, 1107.0860

\bibitem[{{O' Riordan} {et~al.}(2016){O' Riordan}, {Pe'er}, \&
  {McKinney}}]{O'Riordan+16}
{O' Riordan}, M., {Pe'er}, A., \& {McKinney}, J.~C. 2016, \apj, 819, 95,
  1510.08860

\bibitem[{{Pe'er} \& {Casella}(2009)}]{PC09}
{Pe'er}, A., \& {Casella}, P. 2009, \apj, 699, 1919, 0902.2892

\bibitem[{{Pe'er} \& {Markoff}(2012)}]{PM12}
{Pe'er}, A., \& {Markoff}, S. 2012, \apj, 753, 177, 1105.4896

\bibitem[{{Poutanen}(1998)}]{Poutanen98}
{Poutanen}, J. 1998, in Theory of Black Hole Accretion Disks, ed.
  {M.~A.~Abramowicz, G.~Bjornsson, \& J.~E.~Pringle} (Cambridge University
  Press), 100--122, arXiv:astro-ph/9805025

\bibitem[{{Qiao} \& {Liu}(2015)}]{QL15}
{Qiao}, E., \& {Liu}, B.~F. 2015, \mnras, 448, 1099, 1501.03565

\bibitem[{{Remillard} \& {McClintock}(2006)}]{RM06}
{Remillard}, R.~A., \& {McClintock}, J.~E. 2006, \araa, 44, 49,
  arXiv:astro-ph/0606352

\bibitem[{{Ressler} {et~al.}(2015){Ressler}, {Tchekhovskoy}, {Quataert},
  {Chandra}, \& {Gammie}}]{Ressler+15}
{Ressler}, S.~M., {Tchekhovskoy}, A., {Quataert}, E., {Chandra}, M., \&
  {Gammie}, C.~F. 2015, ArXiv e-prints, 1509.04717

\bibitem[{{Russell} {et~al.}(2013){Russell}, {Russell}, {Miller-Jones},
  {O'Brien}, {Soria}, {Sivakoff}, {Slaven-Blair}, {Lewis}, {Markoff}, {Homan},
  {Altamirano}, {Curran}, {Rupen}, {Belloni}, {Cadolle Bel}, {Casella},
  {Corbel}, {Dhawan}, {Fender}, {Gallo}, {Gandhi}, {Heinz}, {K{\"o}rding},
  {Krimm}, {Maitra}, {Migliari}, {Remillard}, {Sarazin}, {Shahbaz}, \&
  {Tudose}}]{Russell+13}
{Russell}, D.~M. {et~al.} 2013, \apjl, 768, L35, 1304.3510

\bibitem[{{Rybicki} \& {Lightman}(1979)}]{RL79}
{Rybicki}, G.~B., \& {Lightman}, A.~P. 1979, {Radiative Processes in
  Astrophysics} (New York, Wiley-Interscience)

\bibitem[{{S{\c a}dowski} {et~al.}(2015){S{\c a}dowski}, {Narayan},
  {Tchekhovskoy}, {Abarca}, {Zhu}, \& {McKinney}}]{Sadowski+15}
{S{\c a}dowski}, A., {Narayan}, R., {Tchekhovskoy}, A., {Abarca}, D., {Zhu},
  Y., \& {McKinney}, J.~C. 2015, \mnras, 447, 49, 1407.4421

\bibitem[{{Sironi} \& {Spitkovsky}(2014)}]{SS14}
{Sironi}, L., \& {Spitkovsky}, A. 2014, \apjl, 783, L21, 1401.5471

\bibitem[{{Tchekhovskoy} \& {McKinney}(2012)}]{TM12}
{Tchekhovskoy}, A., \& {McKinney}, J.~C. 2012, \mnras, 423, L55, 1201.4385

\bibitem[{{Tchekhovskoy} {et~al.}(2012){Tchekhovskoy}, {McKinney}, \&
  {Narayan}}]{TMN12}
{Tchekhovskoy}, A., {McKinney}, J.~C., \& {Narayan}, R. 2012, Journal of
  Physics Conference Series, 372, 012040, 1202.2864

\bibitem[{{Tchekhovskoy} {et~al.}(2010){Tchekhovskoy}, {Narayan}, \&
  {McKinney}}]{TNM10}
{Tchekhovskoy}, A., {Narayan}, R., \& {McKinney}, J.~C. 2010, \apj, 711, 50,
  0911.2228

\bibitem[{{Tchekhovskoy} {et~al.}(2011){Tchekhovskoy}, {Narayan}, \&
  {McKinney}}]{TNM11}
------. 2011, \mnras, 418, L79, 1108.0412

\bibitem[{{Titarchuk}(1994)}]{Titarchuk94}
{Titarchuk}, L. 1994, \apj, 434, 570

\bibitem[{{Yuan} \& {Narayan}(2014)}]{YN14}
{Yuan}, F., \& {Narayan}, R. 2014, \araa, 52, 529, 1401.0586

\bibitem[{{Yuan} {et~al.}(2007){Yuan}, {Zdziarski}, {Xue}, \& {Wu}}]{Yuan+07}
{Yuan}, F., {Zdziarski}, A.~A., {Xue}, Y., \& {Wu}, X.-B. 2007, \apj, 659, 541,
  arXiv:astro-ph/0608552

\end{thebibliography}

\appendix
\numberwithin{equation}{section}
\setcounter{figure}{0}
\section{Effects of Spin on the Redshift Profiles}
\label{sec:redshift_appendix}
\subsection{Analytic Expression for the Redshift Factor}
\label{subsec:circular_motion}
To understand the dependence of the redshift factor on spin, 
we focus on the simple case of circular 
motion in the $r$--$\phi$ plane.
In what follows, we denote quantities in the coordinate (lab) frame
with no primes on the index, in the orthonormal 
``zero angular momentum observer'' (ZAMO) 
frame with one prime, and in the orthonormal fluid frame with two primes. 
The Killing vectors associated with stationarity and axisymmetry are
$\xi^\mu = \delta^\mu_t$, and $\chi^\mu = \delta^\mu_\phi$.
For circular motion, the 4-velocity can be written as
$u^\mu = u^t\left(\xi^\mu + v^\phi \chi^\mu\right)$, where $v^\phi=u^\phi/u^t$.
The condition that the 4-velocity be timelike, 
$g_{\mu\nu}u^\mu u^\nu = -1$, gives
\begin{equation}
    u^t=\left(-g_{tt}-2 g_{t\phi} v^{\phi}-g_{\phi\phi}
        \left(v^{\phi}\right)^{2}\right)^{-1/2}
\end{equation}
Defining $\mathcal{P}_{i}=p_{i}/p_{t}$
, we can write the redshift for circular motion as 
\citep{CB72,CB73,Fanton+97}
\begin{equation}
    \label{eq:redshift_circular}
    \mathcal{R}=\frac{1}{u^{t}\left(1+v^{\phi}\,\mathcal{P}_{\phi}\right)}
\end{equation}
The photon 4-momentum is a null vector and so, in the fluid frame, we have
\begin{equation}
    \mathcal{P}_{r^{\prime\prime}}^2 + \mathcal{P}_{\theta^{\prime\prime}}^2 + 
    \mathcal{P}_{\phi^{\prime\prime}}^2 = 1
\end{equation}
Therefore, $\mathcal{P}_{\phi^{\prime\prime}}$ is bounded by $\pm 1$, 
corresponding to photons emitted in the $\mp\,\phi$ directions.
The ZAMO and fluid frames are simply related by a Lorentz transformation,
and so
\begin{equation}
    \label{eq:lorentz}
    \mathcal{P}_{\phi^{\prime}}=
    \frac{\mathcal{P}_{\phi^{\prime\prime}}-v^{\phi^{\prime}}}
    {1-v^{\phi^{\prime}}\,\mathcal{P}_{\phi^{\prime\prime}}}
\end{equation}
The transformations from the Boyer-Lindquist coordinate basis
to the orthonormal ZAMO basis are given by \citet{Bardeen+72}
\begin{equation}
e_{\nu^\prime} = e^{\mu}{}_{\nu^\prime}\,\partial_\mu,
\qquad e^{\nu^\prime} = e_{\mu}{}^{\nu^\prime}\mathrm{d}x^\mu
\end{equation}
The only non-zero components are
\begin{alignat}{5}
    &e^t{}_{t^\prime}=1/\alpha,& 
    \qquad&e^r{}_{r^\prime}=1/\sqrt{g_{rr}},&
    \qquad&e^\theta{}_{\theta^\prime}=1/\sqrt{g_{\theta\theta}},&
    \qquad&e^\phi{}_{\phi^\prime}=1/\sqrt{g_{\phi\phi}},&
    \qquad&e^\phi{}_{t^\prime}=\Omega/\alpha\\
    &e_t{}^{t^\prime}=\alpha,&
    \qquad&e_r{}^{r^\prime}=\sqrt{g_{rr}},&
    \qquad&e_\theta{}^{\theta^\prime}=\sqrt{g_{\theta\theta}},&
    \qquad&e_\phi{}^{\phi^\prime}=\sqrt{g_{\phi\phi}},&
    \qquad&e_t{}^{\phi^\prime}=-\Omega\,\sqrt{g_{\phi\phi}}
    \end{alignat}
where
\begin{equation}
    \Omega=-\frac{g_{t\phi}}{g_{\phi\phi}},
    \qquad\alpha=\sqrt{-g_{tt}+\Omega^{2}g_{\phi\phi}}
\end{equation}
Transforming from the ZAMO frame to the coordinate frame gives 
\begin{align}
    \label{eq:p_ZAMO}
    \mathcal{P}_{\phi} &=\frac{\sqrt{g_{\phi\phi}}\,\mathcal{P}_{\phi^{\prime}}}
    {\alpha-\Omega\,\sqrt{g_{\phi\phi}}\,\mathcal{P}_{\phi^{\prime}}}\\
    \label{eq:v_ZAMO}
    v^{\phi} &=\frac{\alpha}{\sqrt{g_{\phi\phi}}}\, v^{\phi^{\prime}}+\Omega
\end{align}
Finally, the redshift for circular motion is given by 
equation \eqref{eq:redshift_circular}, with $\mathcal{P}_\phi$ related to 
the fluid frame $\mathcal{P}_{\phi^{\prime\prime}}$ by equations 
\eqref{eq:lorentz} and \eqref{eq:p_ZAMO}, and $v^\phi$ related to the ZAMO 
frame $v^{\phi^\prime}$ by equation \eqref{eq:v_ZAMO}.

\begin{figure*}
    \includegraphics[width=\figfactor\linewidth]{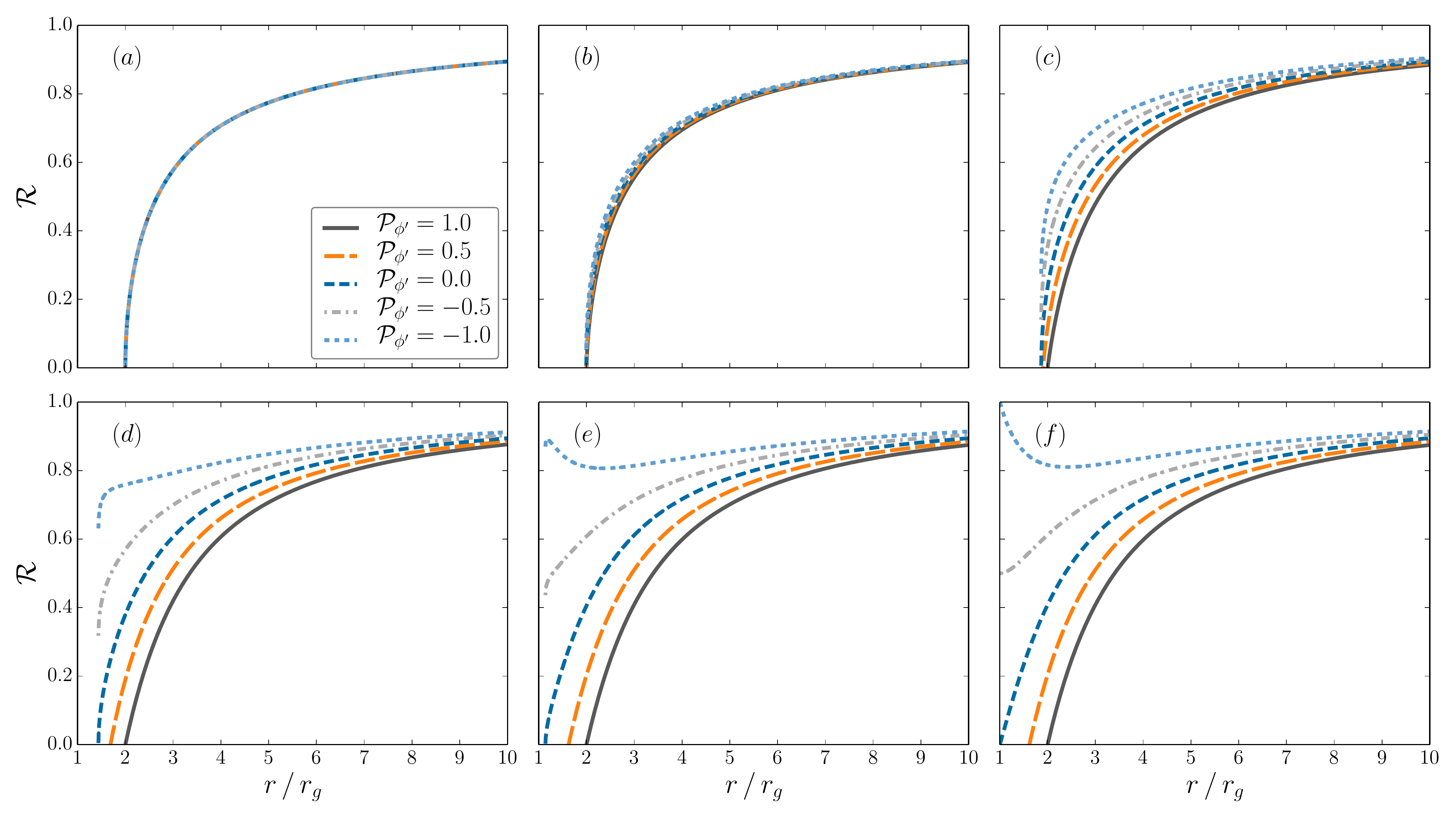}
    \caption{Redshifts for a source with $v^{\phi}=\Omega$ 
        (equation~\ref{eq:redshift_zamo}).
        $(a):\,a=0.1$, $(b):\,a=0.2$, $(c):\,a=0.5$, $(d):\,a=0.9$.
        $(e):\,a=0.99$, $(f):\,a=1.0$. For high 
        spins, photons emitted in the $\phi$ direction in the ZAMO frame suffer
        little redshift until very close to the horizon.} 
    \label{fig:redshifts_zamo_multiple_2}
\end{figure*}

In the special case of a source with zero angular momentum, 
$v^{\phi^{\prime}}=0\Rightarrow v^\phi=\Omega$, and equation 
\eqref{eq:redshift_circular} becomes
\begin{equation}
    \label{eq:redshift_zamo}
    \mathcal{R}=\alpha-\Omega\,\sqrt{g_{\phi\phi}}\,\mathcal{P}_{\phi^{\prime}}
\end{equation}
which is simply the transformation 
$p_t=e_t{}^{\nu^\prime}\,p_{\nu^\prime}$.
In Figure~\ref{fig:redshifts_zamo_multiple_2} we show the redshift for a ZAMO
(equation~\ref{eq:redshift_zamo}), as a function 
of $\mathcal{P}_{\phi^\prime}$, for different spins.
For high spins, photons emitted in the $\phi$ direction 
(those with $\mathcal{P}_{\phi^\prime}= -1$) suffer little redshift until right
outside the horizon. In fact, for a maximally spinning black hole, 
$\mathcal{R}\rightarrow 1$ as $r\rightarrow r_H$.
On average, observers with $\theta=\pi/2$ receive photons with larger 
$\phi$ momentum than observers located at $\theta=0$.
Therefore, observers perpendicular to the spin axis experience a flatter 
redshift profile and so see closer to the horizon.

\subsection{Dependence on the Fluid Velocity}
\label{subsec:model_dependence}
\begin{figure}
    \centering
    \includegraphics[width=\figfactorthree\linewidth]{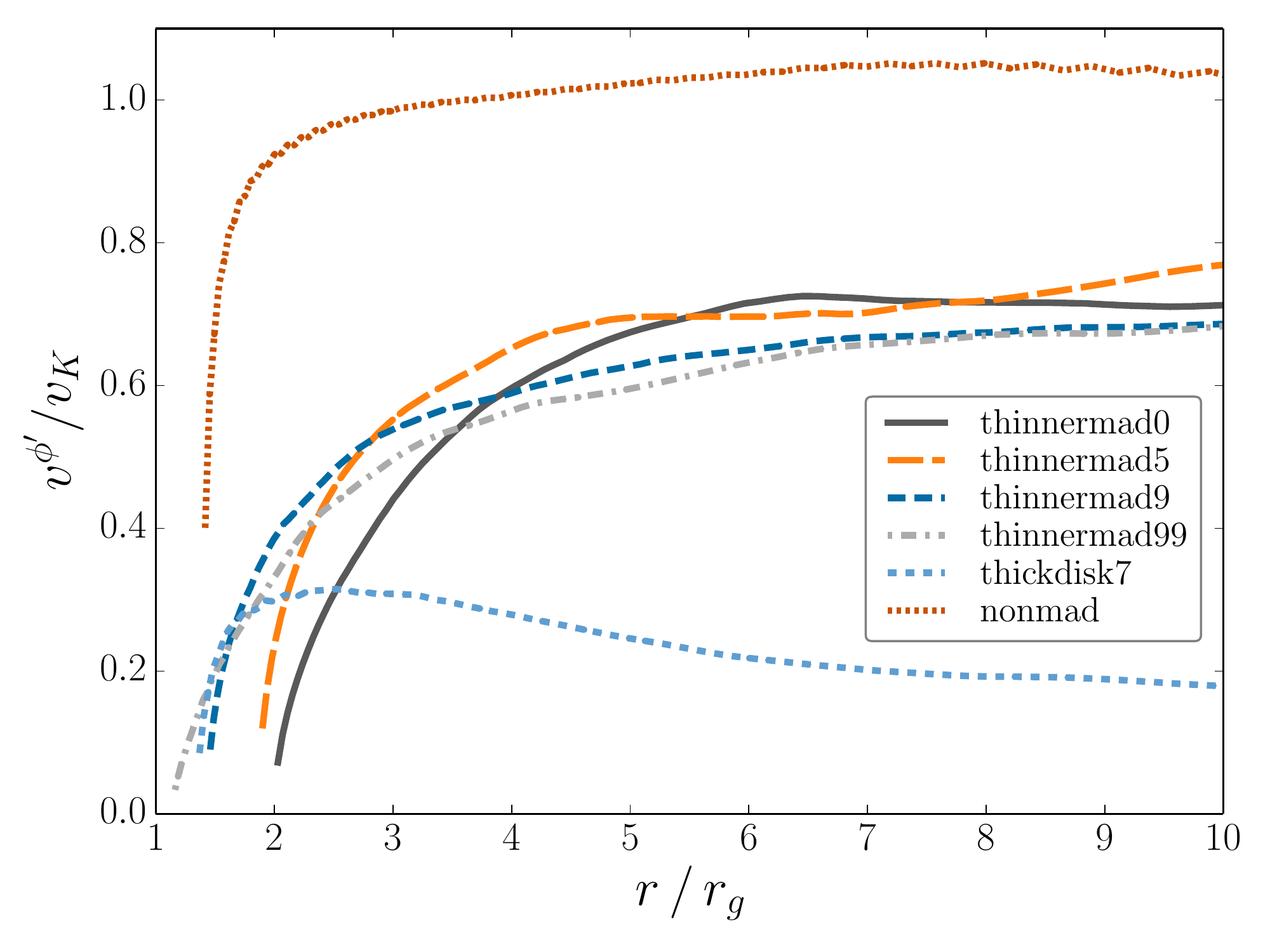}
    \caption{$v^{\phi^\prime}$ in units of $v_K$ for different accretion models.
        The SANE model is roughly Keplerian, while the MAD models are all
        sub-Keplerian. The velocity is approximately a constant fraction of 
        the Keplerian speed until close to the horizon. In the lab frame,
        frame dragging forces the fluid rotate with $v^\phi\rightarrow\Omega_H$ 
        as $r\rightarrow r_H$.    
    }
    \label{fig:vphi_ratio}
\end{figure}

\begin{figure*}
    \includegraphics[width=\figfactor\linewidth]{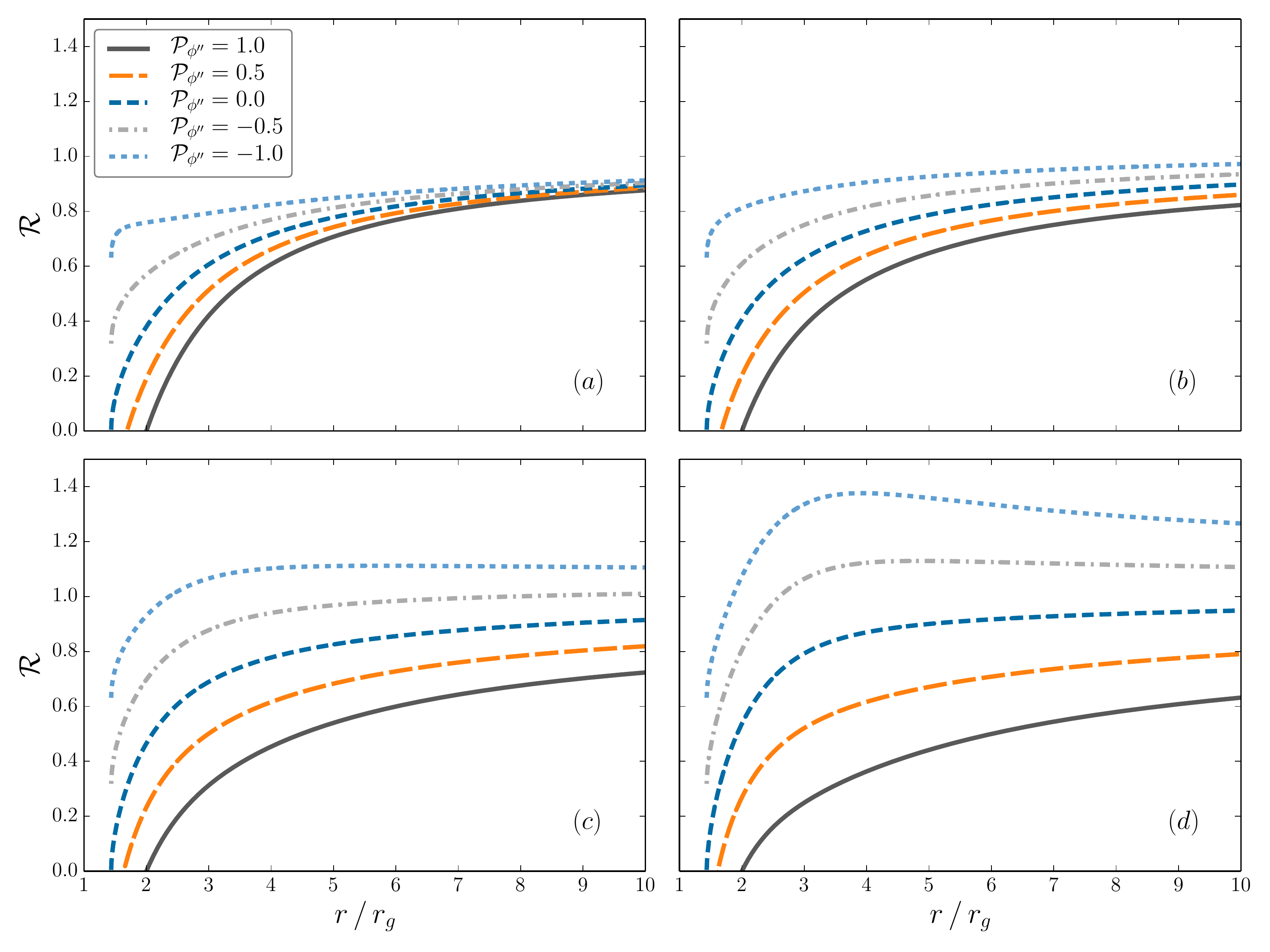}
    \caption{Redshifts for a source with $a=0.9$, and 
        $v^{\phi^\prime}=\epsilon\,v_K$.
        We set $v^{\phi^\prime}=0$ at the horizon, with a transition at     
    $r=r_\text{ISCO}$.
    $(a):\,\epsilon=0.0$, $(b):\,\epsilon=0.2$, $(c):\,\epsilon=0.6$, 
    $(d):\,\epsilon=1.0$.
    } 
    \label{fig:redshifts_circular_motion_multiple_a9}
\end{figure*}

In general, the fluid will have a non-zero $\phi$ velocity in the ZAMO frame.
In Figure~\ref{fig:vphi_ratio} we show $v^{\phi^\prime}$ for different models
from \citet{MTB12}.
We give the velocity in units of the Keplerian speed $v_K = 1/\sqrt{r}$. 
The ``thinnermadx'' models are those considered here 
(where the number x gives the spin), while the 
``thickdisk7'' and ``nonmad'' models are the MAD and SANE models considered 
in \citet{O'Riordan+16}. These models have
$a=0.9375$ and $a=0.92$, respectively. 
In all models, the $\phi$ velocity is approximately a 
constant fraction of the Keplerian speed until close to the horizon (where 
$v^{\phi^\prime}\rightarrow 0$ since $v^{r^\prime}\rightarrow 1$ in this frame).
Motivated by this,
we choose $v^{\phi^\prime}$ to be proportional to $v_K$, and set 
$v^{\phi^\prime}=0$ at the horizon with a smooth transition at $r=r_\text{ISCO}$.
In Figure~\ref{fig:redshifts_circular_motion_multiple_a9} we 
show the effects of varying $v^{\phi^\prime}$ for a black hole with $a=0.9$.
The difference between the observed and comoving power is a factor of 
$\mathcal{R}^2$. Therefore, 
for photons emitted in the $\phi$ direction in the fluid frame, 
the $\phi$ velocity contributes to a maximum factor of $\sim 3$.
Since observed photons will have a spread of 
$\mathcal{P}_{\phi^{\prime\prime}}$, the average difference in power will 
likely be much smaller than this.

\begin{figure}
    \centering
    \includegraphics[width=\figfactorthree\linewidth]{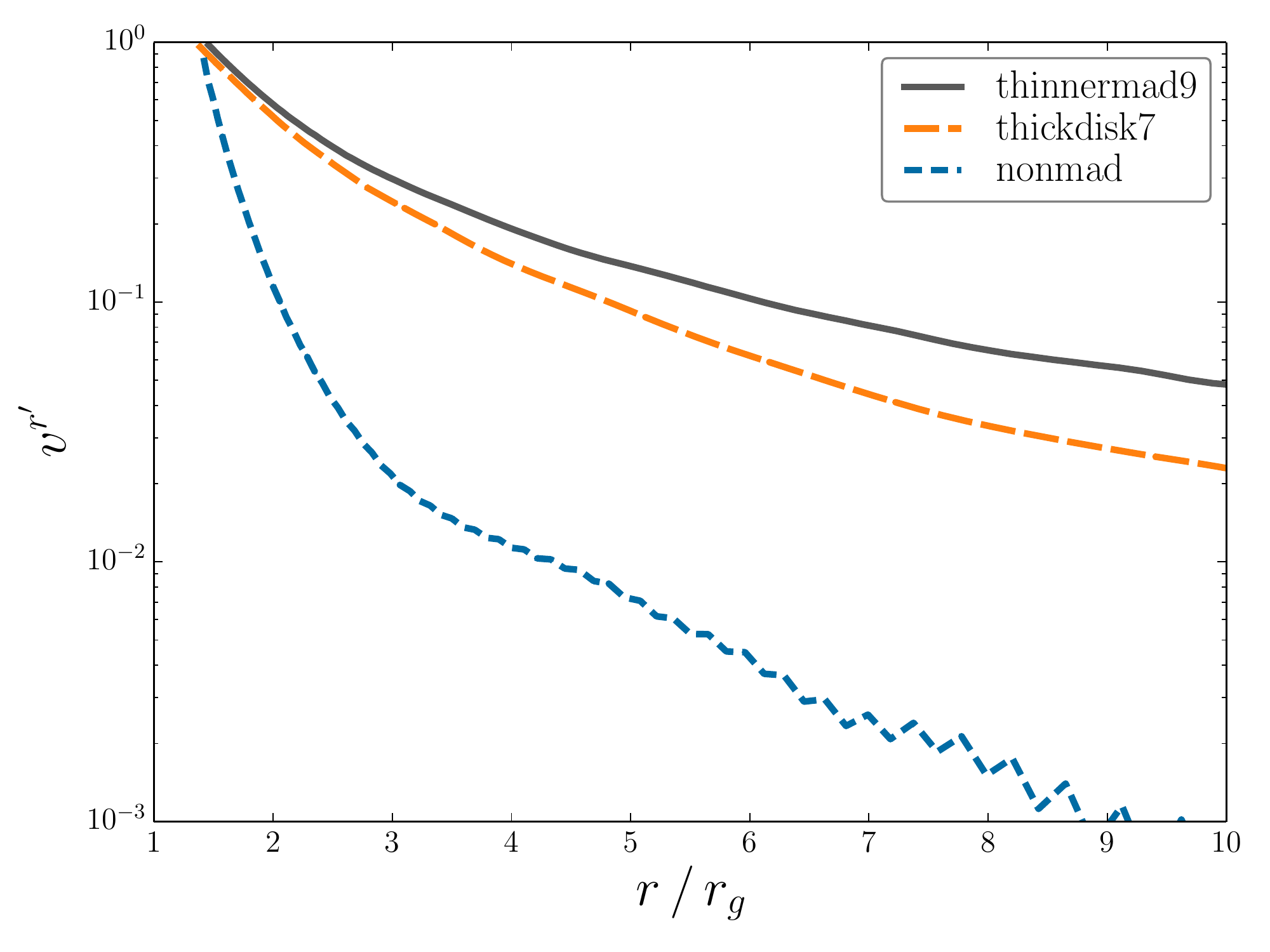}
    \caption{Radial velocity in the ZAMO frame for the MAD and SANE 
        models with similar spin. The radial velocities are comparable in the 
        MAD case, but the velocity profile is significantly different in the 
        SANE case.
    }
    \label{fig:avg_vr}
\end{figure}

\begin{figure}
    \centering
    \includegraphics[width=\figfactorthree\linewidth]{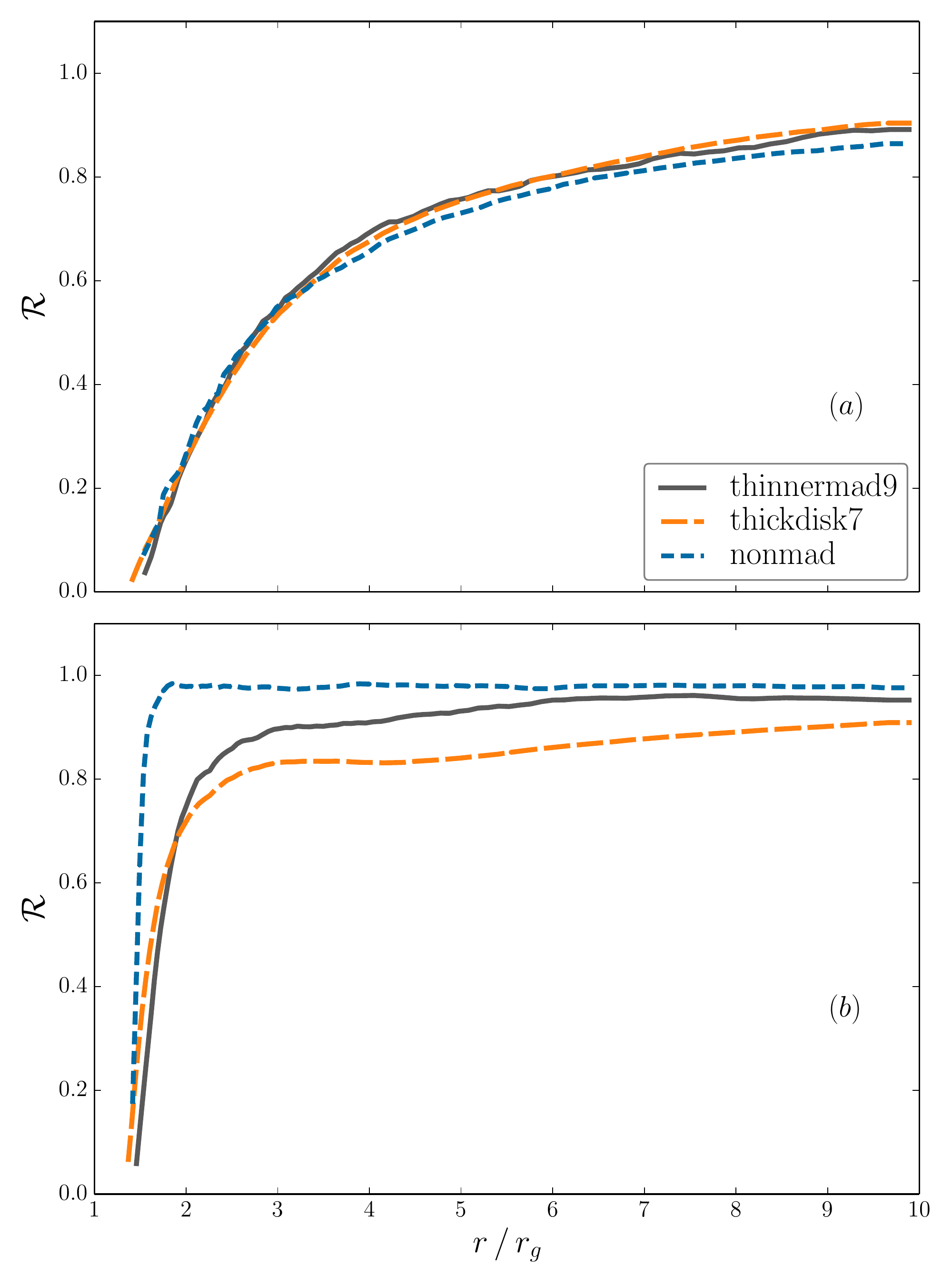}
    \caption{Redshift profiles for different accretion models. The observer is 
        located at $(a):\,\theta=0$, $(b):\,\theta=\pi/2$. 
        Although these models have different velocity fields, 
        the resulting redshift profiles are similar. In particular, for 
        $\theta=\pi/2$ the profiles are reasonably flat until close to the 
        horizon. Therefore, the contributions from spin and viewing angle are
        more important than the model-dependent fluid velocity.
    }
    \label{fig:redshifts_compare}
\end{figure}

To test the sensitivity of the redshift factor to the accretion model, 
we numerically calculate redshift profiles for the ``thinnermad9'', ``thickdisk7'', 
and ``nonmad'' models. These have similar spin, but have different velocity 
fields (see Figure~\ref{fig:vphi_ratio} for the $\phi$ velocity).
In Figure~\ref{fig:avg_vr} we show the radial velocity profiles.
The radial velocities are comparable in the MAD models, 
however these differ significantly from the SANE case.
In Figure~\ref{fig:redshifts_compare} we show the redshift profiles from the
different models, for observers located at $\theta=0$ and $\theta=\pi/2$.
The $\theta=\pi/2$ case should maximize potential deviations. 
Despite differences in the fluid velocity,
the resulting profiles are remarkably similar. In particular, the redshift is
reasonably flat until very close to the horizon.
Therefore, we conclude that the model-dependent velocity contribution to the 
described redshift effect is minor, while the main contributions are the spin 
and viewing angle.

\section{Dependence of the Radiated Power on the Accretion Rate}
\label{sec:accretion_rate}
The GRMHD simulations are scale free, however, introducing radiation forces us 
to specify length, time, and mass/energy scales. 
The length and time scales are set by
the black hole mass $M$. These are the gravitational radius $r_g$, and
light crossing time $t_g=r_g/c$.
Since the fluid mass is $\ll\, M$, we
set the mass/energy scale via the mass accretion rate
\begin{equation}
\label{eq:accretion_rate}
    \dot{M}=\left|\int\mathrm{d}A\,\rho\,u^r\right|
\end{equation}
For some constant $\mu$, we can write this
in terms of the Eddington rate as 
$\dot{M} = \mu\, \dot{M}_\text{Edd}$.
Therefore, for a fixed black hole mass, the rest mass density scales with the 
accretion rate as $\rho\sim\mu$. 
Since energy densities scale in the same sense as $\rho c^2$,
we immediately find $n\sim\mu$, $B\sim\mu^{1/2}$, and $\Theta\sim\mu^0$.
The last relation follows from the fact that, for a perfect fluid,
$\Theta\sim u/\rho c^2$, where $u$ is the internal energy density.
Finally, the synchrotron emissivity scales with the accretion rate as
$nB^2\Theta^2\sim\mu^2$, and the synchrotron frequency scales as
$B\Theta^2\sim\mu^{1/2}$.

\end{document}